\documentclass[]{spie}

\usepackage[]{graphicx}
\usepackage{amsmath, amssymb, amsfonts}
\usepackage[english]{babel}
\usepackage[]{epstopdf}
\usepackage{booktabs,fixltx2e}
\usepackage[flushleft]{threeparttable}

\usepackage{wrapfig}

\usepackage{hyperref}

\title{Capabilities of ACAD-OSM, an active method for the correction of aperture discontinuities} 

\author{Johan Mazoyer\supit{a,b}, Laurent Pueyo\supit{a}, Mamadou N'Diaye\supit{c}, Kevin Fogarty\supit{b}, Lucie Leboulleux\supit{a,d,e} Sylvain Egron\supit{a,d,e}, Colin Norman\supit{b}
\skiplinehalf
\supit{a} Space telescope Science Institute, 3700 San Martin Drive, Baltimore, MD 21218, USA\\
\supit{b} Department of Physics and Astronomy, Johns Hopkins University, Baltimore, MD, USA\\
\supit{c} Universit\'e C\^ote d'Azur, Observatoire de la C\^ote d'Azur, CNRS, Laboratoire Lagrange, Bd de l'Observatoire, CS 34229, 06304 Nice cedex 4, France\\
\supit{d} Aix Marseille Universit\'e, CNRS, Laboratoire d'Astrophysique de Marseille, UMR 7326, 13388, Marseille, France \\
\supit{e} Office National d'Etudes et de Recherches A\'erospatiales, 29 Avenue de la Division Leclerc, 92320 Ch\^atillon, France
}

\authorinfo{Further author information: contact Johan Mazoyer at jmazoyer@jhu.edu}
 
\begin{document} 
  
\maketitle 

\begin{abstract}
The increasing complexity of the aperture geometry of the future space- and ground based-telescopes will limit the performance of the next generation of coronagraphic instruments for high contrast imaging of exoplanets. We propose here a new closed-loop optimization technique using two deformable mirrors to correct for the effects of complex apertures on coronagraph performance, alternative to the ACAD technique previously developed by our group. This technique, ACAD-OSM, allows the use of any coronagraphs designed for continuous apertures, with complex, segmented, apertures, maintaining high performance in contrast and throughput.
We show the capabilities of this technique on several pupil geometries (segmented LUVOIR type aperture, WFIRST, ELTs) for which we obtained high contrast levels with several deformable mirror setups (size, number of actuators, separation between them), coronagraphs (apodized pupil Lyot and vortex coronagraphs) and spectral bandwidths, which will help us present recommendations for future coronagraphic instruments. We show that this active technique handles, without any revision to the algorithm, changing or unknown optical aberrations or discontinuities in the pupil, including optical design misalignments, missing segments and phase errors.

\end{abstract}



\section{Introduction}
\label{sec:intro}

The next generation of exoplanet imagers on the ground with Extremely Large Telescopes (ELTs) \cite{macintosh06, kasper08,davies10,quanz15c} and in space with WFIRST \cite{spergel15} will seek $10^{-9}$ contrasts to find older Jupiter-like planets. Further along, only envisioned space telescopes such as LUVOIR \cite{dalcanton15} or HabEx \cite{mennesson16} will aim for the $10^{-10}$ contrast required to detect Earth like planets around nearby stars and investigate the presence of bio-markers on their surface.

As the telescope primary mirror becomes larger, the complexity of the aperture geometry increases. First, the central obscuration can reach up to 36\% of the pupil diameter as in the case of WFIRST. Second, the support structure becomes thicker as the the telescope secondary mirror becomes larger, blocking more light and producing stronger diffractive effects in the star Point Spread Function (PSF). Finally, the segmentation of the aperture, already present on the Keck Observatory, will increase with up to hundreds of segments for the ELTs. Following JWST with 18 primary mirror segments, the segmentation trend is likely to increase in space. These complex apertures will have a strong impact on the coronagraph performance.

In this proceeding, we introduce a new active technique to correct for aperture discontinuities, the Active Compensation of Aperture Discontinuties-Optimized Stroke Minimization (ACAD-OSM). In Section~\ref{sec:acad}, we recall the previous ACAD method \cite{pueyo13} and present our new adaptive interaction matrix algorithm for the active correction of aperture discontinuities using two DMs. We then show the advantage of this technique over fixed apodization to correct for evolution and unknowns in the system (segment failure, phase aberrations and misalignments).

\section{The ACAD technique: initial goals and evolution}
\label{sec:acad}

\begin{table*}
\centering
\caption{This table shows the parameters of the coronagraphs used in this article for different apertures.}
\label{tab:ularasa}
\begin{tabular}{|c|c|c|c|c|}
\hline
              & FPM              & LS inner radius &  LS outer radius & Optimized for               \\ \hline  \hline
\multicolumn{5}{|c|}{36\% central obscuration (WFIRST)}                                                                                                                                                               \\ \hline  \hline

PAVC6 & Charge 6 vortex    & 55\%         & 100\%              & ideal                                                \\ \hline  \hline
\multicolumn{5}{|c|}{17\% central obscuration (SCDA)}                                                                                                                                                                 \\ \hline  \hline
APLC          & Opaque Lyot mask: 4 $\lambda_0/D_{ap}$ radius & 30\%   & 92\%               & \begin{tabular}[c]{@{}c@{}}c = $10^{-10}$ (10\% BW)\end{tabular}             \\ \hline

PAVC6 & Charge 6 vortex    & 41\%        & 100\%            & ideal                                                                       \\ \hline \hline
\multicolumn{5}{|c|}{30\% central obscuration (E-ELT)}                                                                                                                                                                 \\ \hline  \hline
PAVC6 & Phase charge 6 vortex    & 39\%      & 100\%              & ideal    \\ \hline \hline
\multicolumn{5}{|c|}{14\% central obscuration (LUVOIR)}                                                                                                                                                                 \\ \hline  \hline
PAVC6 & Phase charge 6 vortex    & 28\%      & 100\%              & ideal                                                                   \\ \hline

\end{tabular}
\end{table*}

\subsection{Description of two DM aperture correcting technique goals}
\label{sec:acad_desc}

\begin{figure}
\begin{center}
 \includegraphics[width = .48\textwidth]{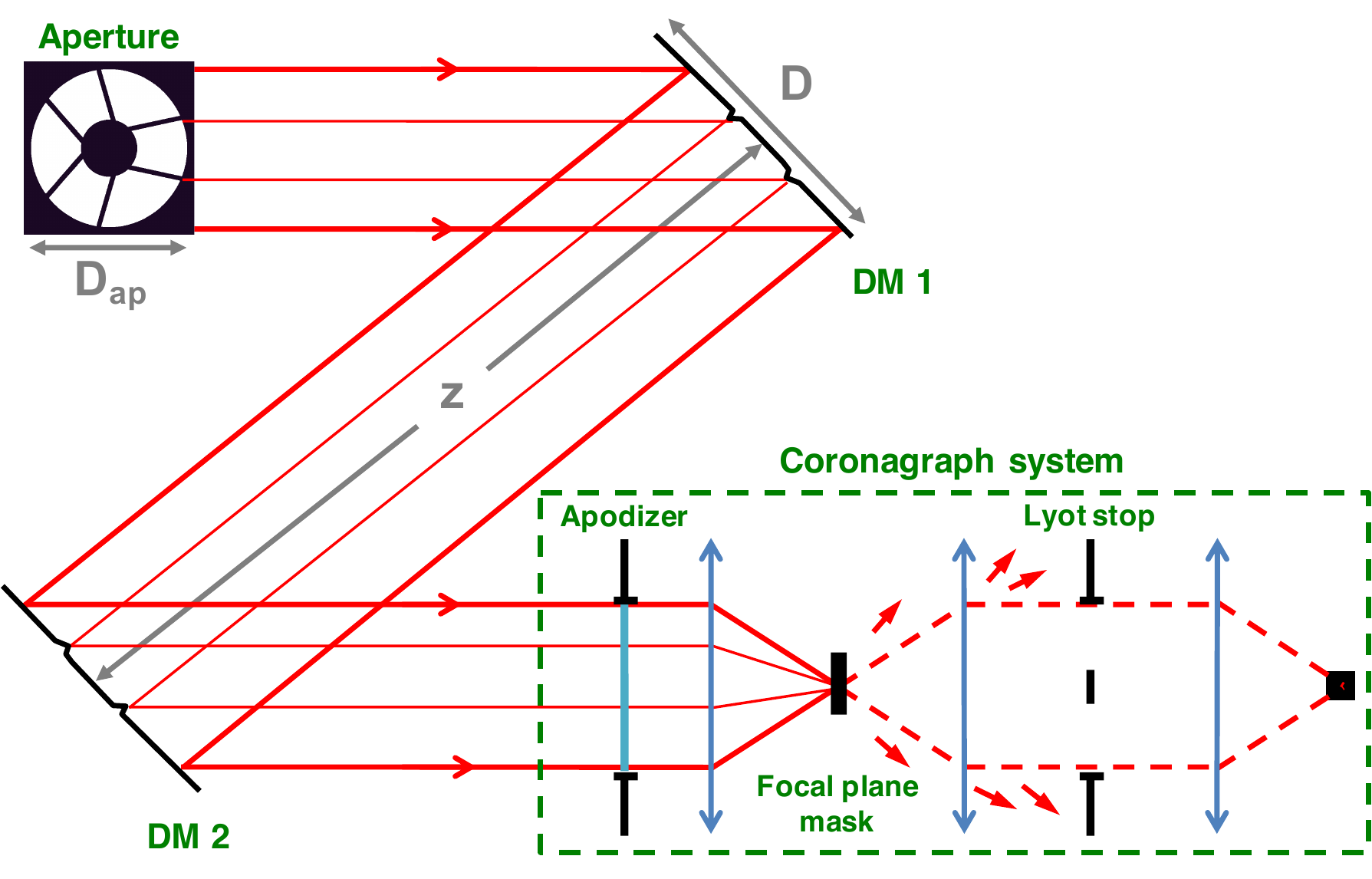}
 \end{center}
\caption[plop]
{\label{fig:schema_ACAD} Schematic representation of a two DM system and a coronagraph. We show the distances z, D and $D_{ap}$ on this optical layout.}
\end{figure}

Figure~\ref{fig:schema_ACAD} shows a schematic representation of the optical design used in this proceeding. Before reaching the first DM, light from a target star is reflected by the primary mirror onto the secondary mirror of the telescope. The resulting pattern, showing all the light blocked both by the primary (segmentation) and by the secondary mirror (central obscuration) and its structures, at the entrance of the instrument is called the telescope aperture. We consider a circular aperture with a diameter $D_{ap}$, although this geometry is not a requirement for the technique described here. We also assume null phase and amplitude aberrations and only focus on the problem of correcting for discontinuities of the aperture, except in Section~\ref{sec:phase_errors}. 

The beam is then reflected on the first DM that is located in a pupil plane. The two DMs are assumed to be square with the same size $D\times D$, and the same number of actuators $N_{act}\times N_{act}$. We also assume that the DM is larger than the aperture: $(1+\alpha)D_{ap} = D$ with $\alpha = 10\%$. We will often write the size $D$ in the form $N_{act} *$ IAP, where IAP is the DM inter-actuator pitch. The beam is propagated over a distance $z$ to the second DM that is found outside of the pupil plane. To avoid beam walks outside the second DM at large separations, in this proceeding, we assume that the region outside the second DM is reflective but not actuated. After both reflections, we re-image a pupil plane in which the coronagraph entrance pupil takes place with an apodizer. Downstream this part, we have a coronagraph design with a focal plane mask (FPM), a Lyot stop (LS) in the relayed pupil plane, and finally a detector in the final image plane.

In this proceedings, the two coronagraph designs are optimized to correct for the effects of the pupil with central obscuration and the active correction is only used to address the other aperture discontinuities (segmentation and secondary struts). The first design is the PAVC \cite{fogarty17} using a vortex FPM and an apodization. We simulate the vortex coronagraph using the the numerical technique described in \cite{mazoyer15}. As a second design, we will use the APLC \cite{soummer03,soummer11, ndiaye16} using an opaque FPM. This concept is therefore not able to reach small separations but it is very robust to small jitter residuals and resolved stars. The apodization have been optimized to maximize the throughput, while providing a $10^{-10}$ contrast over a 10$\%$ BW. Table~\ref{tab:ularasa} summarizes the parameters of the coronagraphs optimized for central obscuration used in this proceeding. In the following, $\lambda_0$ and $\Delta \lambda$ are the central wavelength and width of the considered spectral BW. The central wavelength $\lambda_0$ is set at 550 nm. 

\subsection{The ray optic solution (ACAD-ROS)}
\label{sec:acad_ros}

In previous papers \cite{pueyo13, pueyo14, mazoyer15, mazoyer16b}, the ACAD method was described as a two step method: (i) an semi-analytic ray optic solution (hereafter ACAD-ROS) for the DM shapes obtained by solving the Monge-Amp\`ere equation, (ii) a SM algorithm \cite{pueyo09} to minimize the strokes applied to the DM actuators to achieve a given contrast in the DH.

However, this technique has several important limitations. First, as showed in Section~\ref{sec:compar_OSM_ROS}, this method gives limited results in contrast in challenging cases such as a large BW ($\Delta \lambda /\lambda_0 > 10$ \%) or large struts \cite{pueyo13, mazoyer16b}. Secondly, the DM strokes introduced in the ACAD-ROS step are proportional to the Fresnel number $D^2/\lambda z$ and range from a few hundreds of nanometers to a few micrometers. Using the HiCAT setup, we previously showed that the required ACAD-ROS solution strokes could barely be applied on Boston Micromachines (BMC\footnote{\url{http://www.bostonmicromachines.com/deformable-mirrors.html}}) DMs \cite{mazoyer16b}. More importantly, these very large DM actuator strokes from ACAD-ROS tend to scatter the light of the off-axis PSF, leading to an important effect on the throughput of a companion at large separation. Such a scattering of the PSF at large separation due to high strokes on the mirrors is one of the main limitations of static mirror apodization \cite{guyon2005}. Finally, because ACAD-ROS flatten semi-analytically the wavefront in the pupil plane, one practical disadvantage of this technique is that the solution cannot be applied to the DMs by only using a focal plane wavefront sensing technique.

\subsection{Original SM algorithm }
\label{sec:strokemin}

The SM algorithm \cite{pueyo09} is a correction technique for high-contrast imaging that can be used with single or several DM designs. It first uses a technique of estimation of the speckle complex electrical field in focal plane to build an interaction matrix. This matrix links DM movements to their effect in the speckle complex electrical field estimation. For the detailed construction of an interaction matrix, see \cite{mazoyer13c}. This matrix is then used to find the relevant DM shape(s) to correct for any speckle field due to phase and amplitude errors and produce a DH in the focal plane of a coronagraph. Like most correction algorithms, this method relies on an assumption of small aberrations relative to $\lambda$ to ensure linearity. This linear algorithm does not minimize the contrast but the strokes introduced on a DM to reach a target contrast $C_{target}$. To find a local minimum in contrast with a small amount of strokes, we lower the target focal plane contrast at each iteration $k$ by the gain $\gamma$ of the algorithm :
\begin{equation}
\label{eq:gainSM}
C_{target}[k+1] = (1-\gamma) C_{target}[k]\,\,\,\,\, 0<\gamma<1\,\,.
\end{equation}

\begin{figure}
\begin{center}
 \includegraphics[width = .48\textwidth, trim=0.1cm 4.5cm 4.5cm 4cm, clip = true]{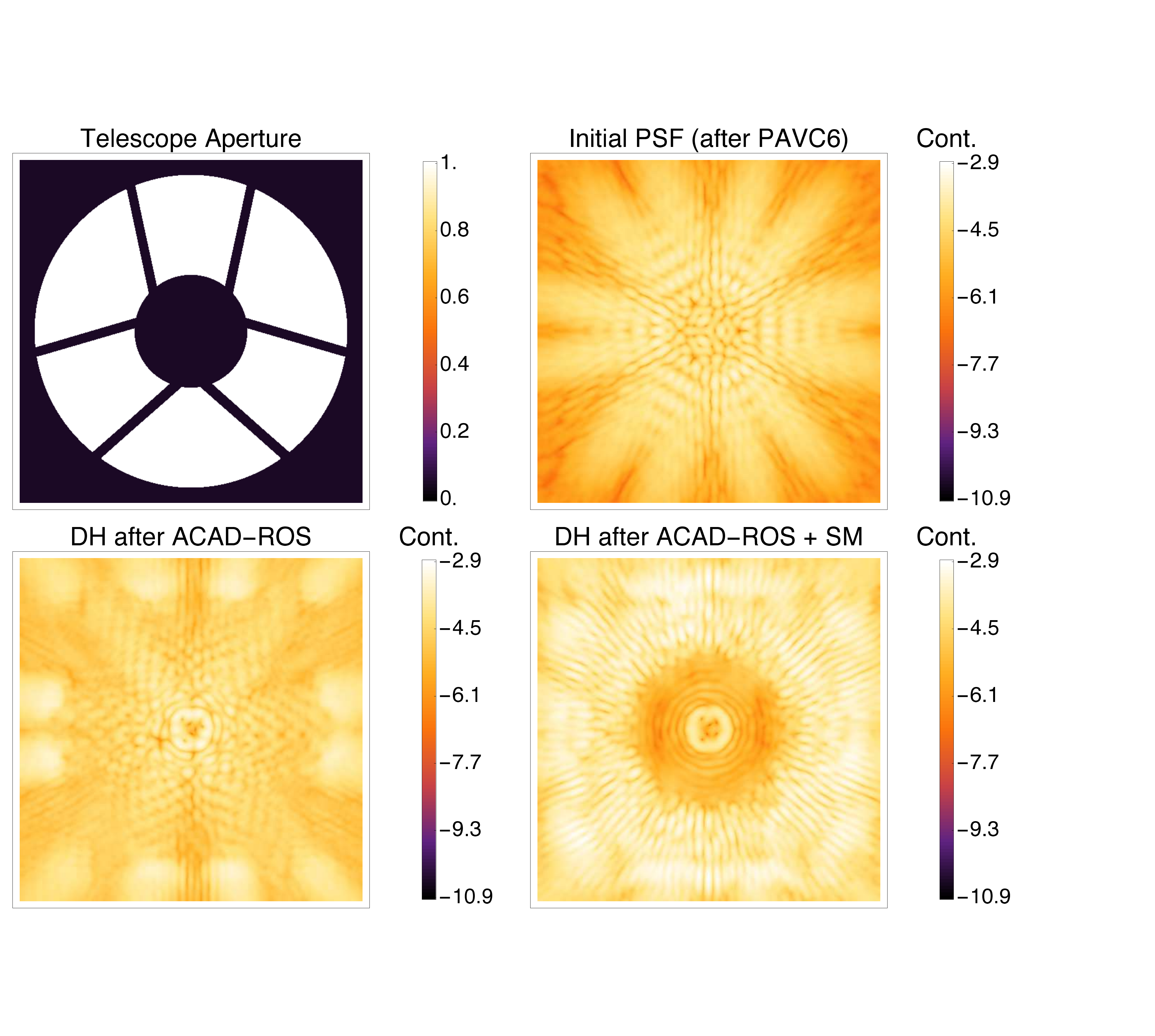}
 \includegraphics[width = .48\textwidth, trim= 0.1cm 4.5cm 4.5cm 4cm, clip = true]{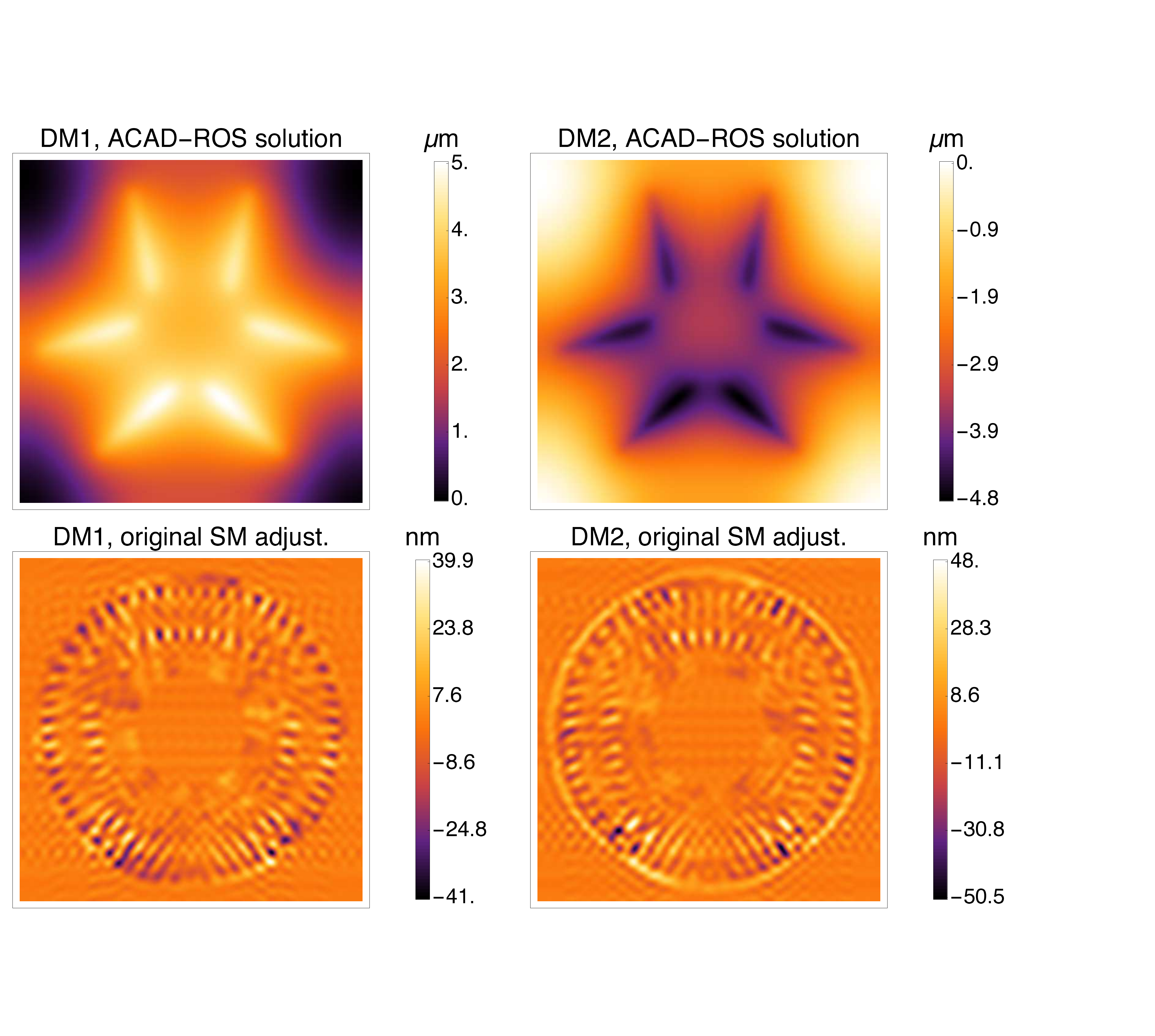}
 \end{center}
\caption[fig:wfirst_dhdm_ACADROS]
{\label{fig:wfirst_dhdm_ACADROS} Results of the ACAD-ROS + SM method (WFIRST aperture, charge 6 PAVC, $N_{act} = 48$, IAP = 1 mm, D = $48 * 1$ mm, $z = 1$ m, $\Delta \lambda /\lambda_0 = $ 10\% BW). \textbf{Left:} Evolution of the PSF at each step. Top, left: WFIRST aperture. Top right: the initial PSF after the coronagraph. Bottom, left: the DH created after the ACAD-ROS solution. Bottom, right: the final 3-10 $\lambda_0/D_{ap}$ DH created after ACAD-ROS and SM algorithms. \textbf{Right:} DM shapes at each step. Top: the two ACAD-ROS DM shapes. Bottom: DM shape adjustments after the SM step.}
\end{figure}

The performance for this aperture with ACAD-ROS + SM are shown in Fig.~\ref{fig:wfirst_dhdm_ACADROS}.The initial state of the PSF after this aperture (with flat DM shapes) is shown in Fig.~\ref{fig:wfirst_dhdm_ACADROS} (left) top right panel. First, the two DM shapes are analytically retrieved using ACAD-ROS method (Fig.~\ref{fig:wfirst_dhdm_ACADROS}, right, top panels). These DM shapes with 5~$\mu$m peak-to-valley strokes are remapping the electrical field to average the effect of the struts inside the DH (Fig.~\ref{fig:wfirst_dhdm_ACADROS}, left, bottom left panel). Finally, we apply the SM algorithm to create DM adjustments of a few tens of nanometers (Fig.~\ref{fig:wfirst_dhdm_ACADROS}, right bottom panels) on top of the ACAD-ROS shapes and generate the a 3-10 $\lambda_0/D_{ap}$ DH in the focal plane image, see Fig.~\ref{fig:wfirst_dhdm_ACADROS}, left, bottom right panel.

\newpage
\subsection{The new adaptive interaction matrix method: ACAD-OSM algorithm}
\label{sec:acad_osm_desc}

\begin{wrapfigure}{l}{0.25\textwidth}
  \begin{center}
    \includegraphics[width=0.22\textwidth]{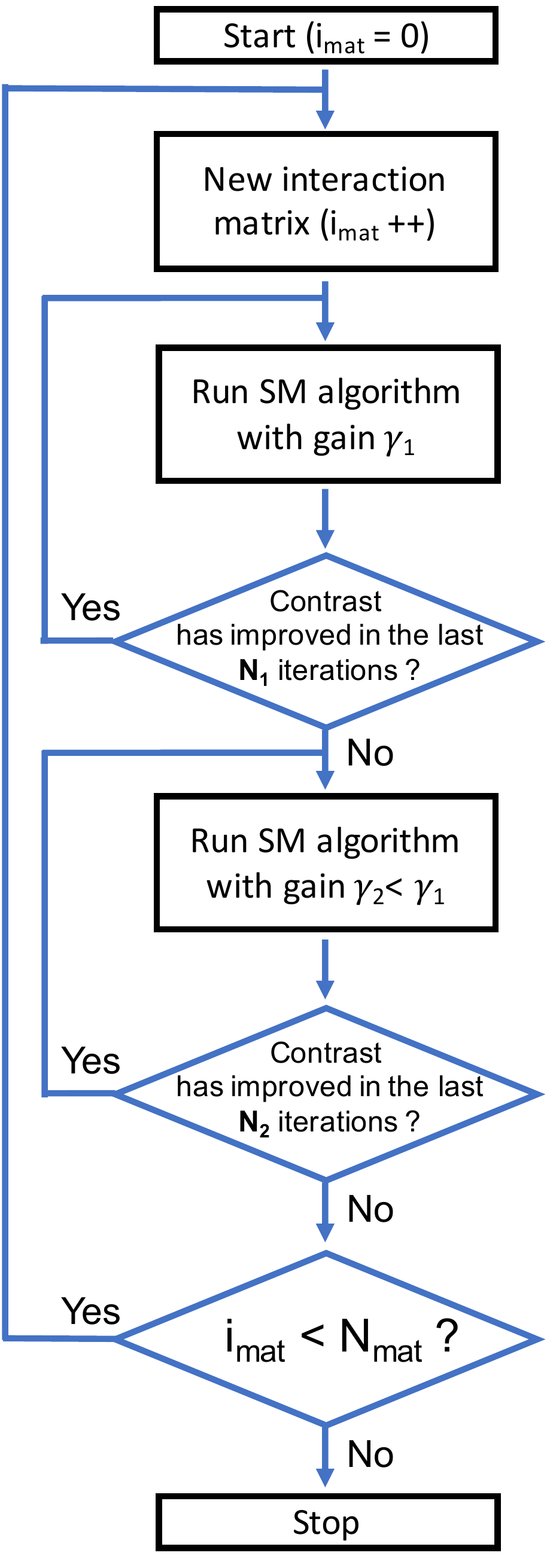}
  \end{center}
  \caption[block_diagram_acad] 
 {\label{fig:block_diagram_acad}Block diagram of the ACAD-OSM algorithm.}
 \vspace{-1.cm}
\end{wrapfigure}
We introduce a new active technique for the correction of aperture discontinuities: ACAD-OSM. This technique, as most correcting technique for small phase and amplitude aberrations, require a focal plane wavefront sensing technique. Several wavefront sensing techniques have been developed over the past two decade, either using small hardware modifications \cite{baudoz06, ndiaye16b} or by introducing known phase patterns in the pupil with the DMs \cite{borde06,giveon07,paul13,riggs16} to measure these errors. Most of these techniques have been tested experimentally in the presence of clear, circular apertures \cite{trauger07,mazoyer14}. We assume here a perfect sensing method to estimate the complex electrical field in focal plane.

The SM algorithm shows good performance for the correction of small phase errors but it diverges quickly for the correction of aperture discontinuities. This behavior was expected since the required strokes for the correction of aperture discontinuities are too large to assume a linear relation between the DM movements and their effects in the focal plane. Indeed, aperture discontinuity usually requires correction with larger strokes than the phase and amplitude errors.

To avoid this problem, we build an optimized stroke minimization algorithm, shown in the form of a block diagram in Fig~\ref{fig:block_diagram_acad}. We use a variable SM loop gain $\gamma$ (as define in Eq~\ref{eq:gainSM}). For the first iterations in the linear range of the initial DM shapes, we set the gain high enough $\gamma = \gamma_1$ to converge faster. When the algorithm starts to diverge or oscillate, the DM shapes derived from the linear SM algorithm do not improve the DH contrast anymore. Consequently, after $N_1$ iterations without improvement, we decrease the SM gain in an attempt to dig the DH forward: $\gamma = \gamma_2$. Secondly, if the contrast still does not improve with the new gain after $N_2$ iterations, we consider that the limit of the linearity range allowed by the initial DM shapes has been reached. 

\begin{figure}
\begin{center}
 \includegraphics[width = .48\textwidth, trim= 0.1cm 4.5cm 4.5cm 4cm, clip = true]{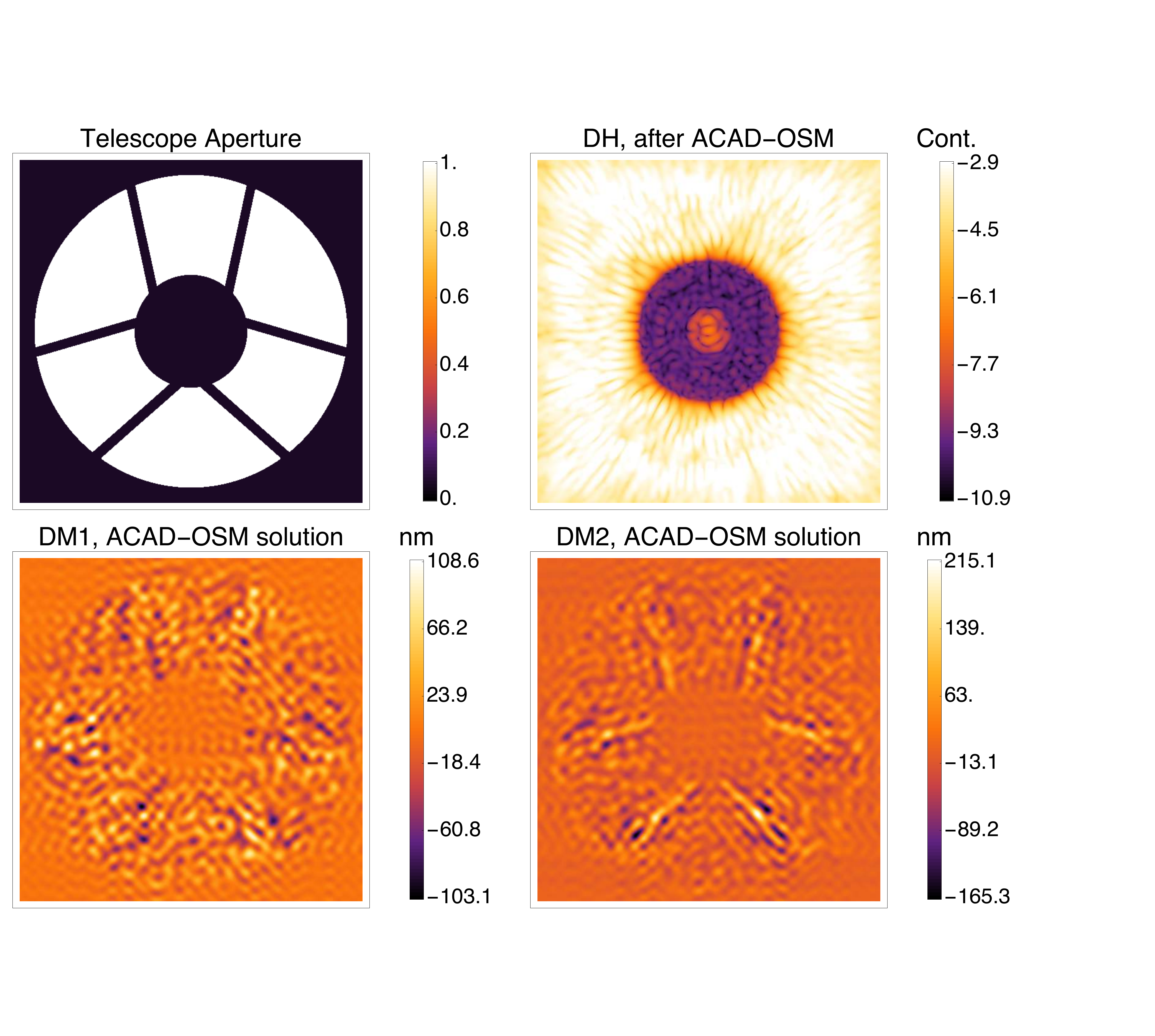}
 \end{center}
\caption[fig:wfirst_dh_ACADOSM]
{\label{fig:wfirst_dh_ACADOSM} WFIRST aperture with the ACAD-OSM technique (charge 6 PAVC, $N_{act} = 48$, IAP = 1 mm, D = $48 * 1$ mm, $z = 1$ m, $\Delta \lambda /\lambda_0 = $ 10\% BW). Top left: WFIRST aperture. Top right: the final 3-10 $\lambda_0/D_{ap}$ DH. Bottom: the DM shapes obtained using ACAD-OSM.}
\end{figure}

We then recenter this range around the final DM shapes from the previous step. These shapes are set as the new reference level to build a new interaction matrix. We repeat this operation $N_{mat}$ times. With every matrix re-computation, the contrast improvement tends to decrease, implying that the limitation of the convergence is not anymore the linearity range but the proximity of a local minimum. We typically stop after the eighth matrix ($N_{mat} = 8$), since in most cases the additional contrast improvement is less than a factor of 2. We use $N_1 =10$, $N_2 =20$, $\gamma_1 = 5\%$ and $\gamma_2 = 2\%$ in this article. In Fig~\ref{fig:wfirst_dh_ACADOSM}, we show the result of ACAD-OSM with a Charge 6 PAVC with the WFIRST aperture, with $M_{mat} = 8$ consecutive interaction matrices.

\subsection{Comparison between the two algorithms}
\label{sec:compar_OSM_ROS}

In this proceeding, we define the contrast as the amount of star light at a given point of the focal plane normalized to the intensity peak of the on-axis PSF in the final focal plane and in the absence of the coronagraph FPM. The throughput is defined as follow: for an off-axis source at a given separation from the star, this value represents the ratio of the energy reaches the off-axis PSF core in the final image plane (inside a photometric aperture of 0.7 $\lambda_0/D_{ap}$ radius centered around the expected position of the PSF) to the energy in the on-axis PSF core (inside the same photometric aperture) when the 2 DM and the coronagraph systems are removed. 

\begin{figure}
\begin{center}
 \includegraphics[width = .48\textwidth, trim= 1.1cm 0.8cm 0.7cm 0.3cm, clip = true]{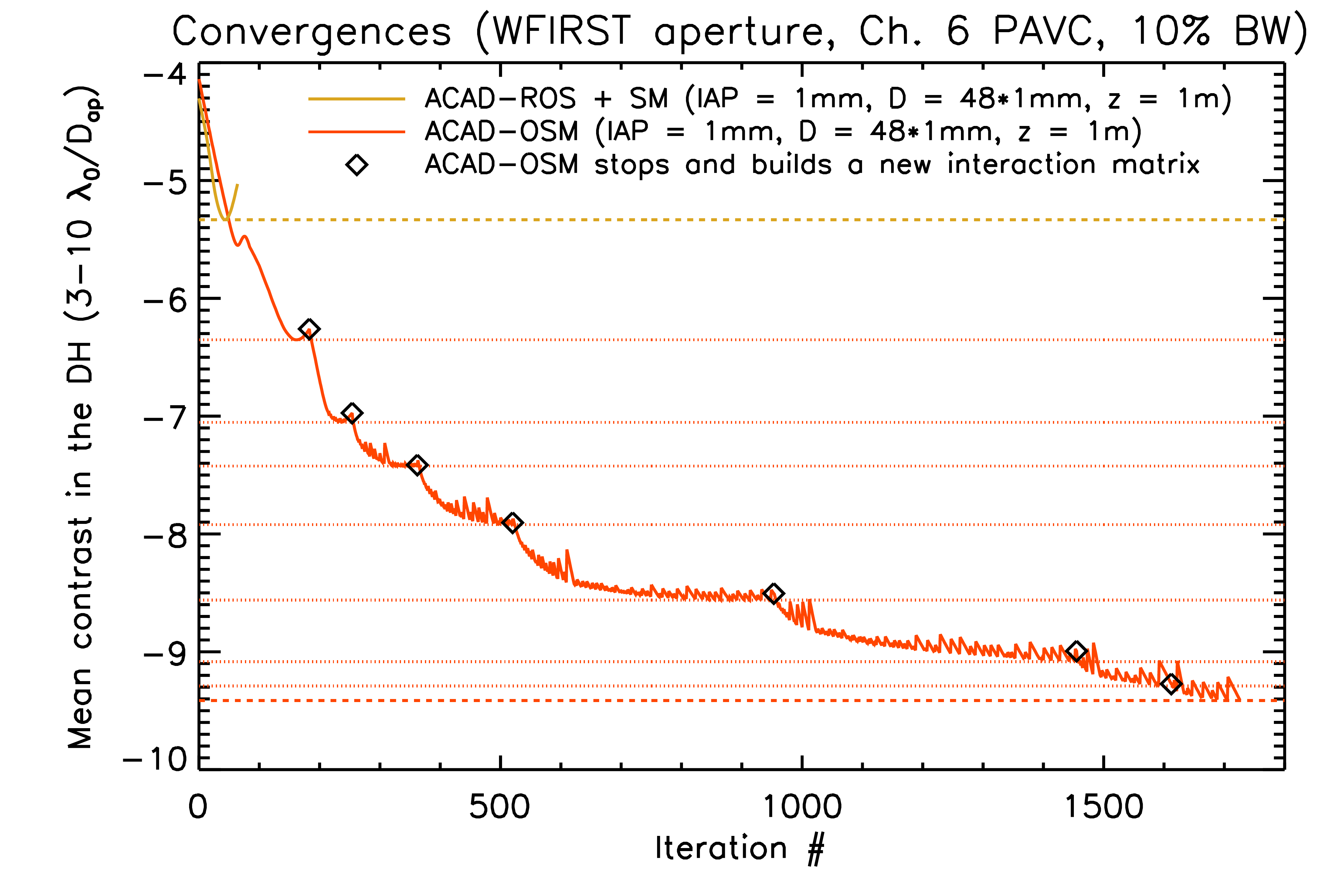}
 \includegraphics[width = .48\textwidth, trim= 1cm 0.8cm 0.7cm 0.3cm, clip = true]{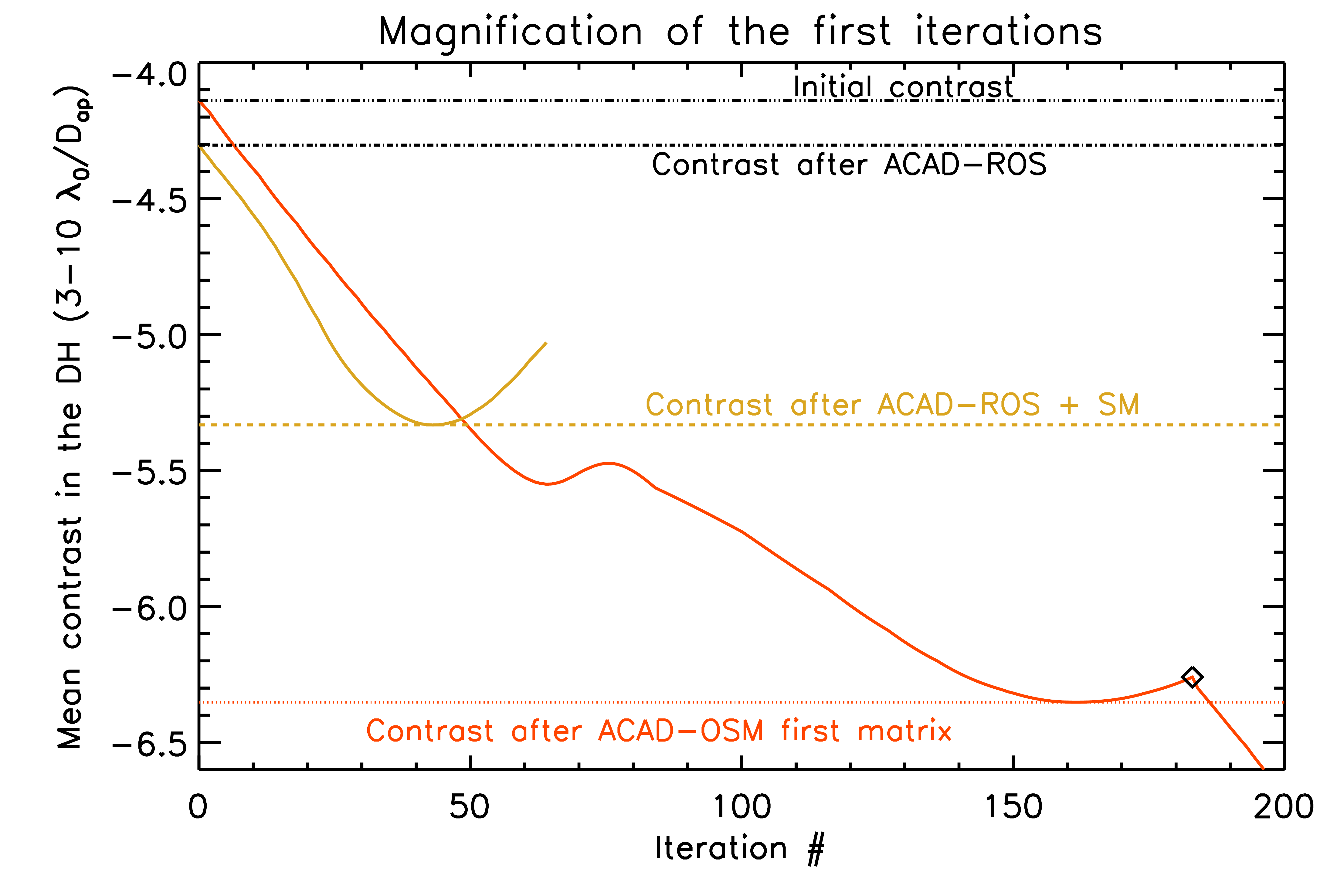}
 \end{center}
\caption[fig:Iter_wfirst]
{\label{fig:Iter_wfirst} Convergence of the mean contrast level in the DH as a function of the number of iterations for the ACAD-ROS + SM solution (yellow curve) and for the ACAD-OSM solution in 8 matrices (red curve). We used a charge 6 PAVC coronagraph and the DM setup is the one used in the WFIRST mission: WFIRST aperture, $N_{act} = 48$, IAP = 1 mm, D = $48*1$ mm, $z = 1$ m, $\Delta \lambda /\lambda_0 = $ 10\% BW. The black diamonds are located on the iterations where the algorithm stops and recalibrates with a new interaction matrix. At each step, we indicate the contrast level by a horizontal dotted line. Final contrasts for both methods are shown with dash lines. On the bottom plot, we show a magnification of the first 200 iterations from the top plot.}
\end{figure}

Fig.~\ref{fig:Iter_wfirst} left plot shows the mean contrast level in the 3-10 $\lambda_0/D_{ap}$ DH as a function of the iteration number with ACAD-ROS+SM (in yellow) and ACAD-OSM (in red) methods. For both cases, we consider the WFIRST aperture, a Charge 6 PAVC, and the WFIRST DM setup: DM size D $48\times0.3$ mm, number of actuators ($48\times48$ actuators) and inter-DM distances $z$. We correct for a 10\% BW centered around $\lambda_0 = 550$ nm, and with no phase aberrations. For ACAD-OSM, the 8 black diamonds show the iterations at which the algorithm stops and builds a new interaction matrix. In this configuration, the red horizontal dotted lines represent the best contrast achieved with each interaction matrix. Fig.~\ref{fig:Iter_wfirst} right plot shows a magnification of the first 200 iterations. We plot black horizontal lines to show the initial contrast and the contrast level after the ACAD-ROS technique only. The thicker dashed lines represent the final contrast: yellow after ACAD-ROS and SM and red for ACAD-OSM. 

The initial DH contrast level with initial flat DM shapes and after ACAD-ROS is $10^{-4.14}$ and $10^{-4.3}$, showing an improvement of only 1.4. In the first 50 iterations, the ACAD-ROS and SM method shows better performance than the ACAD-OSM. However, the SM algorithm after ACAD-ROS diverges after 40 iterations, due to the fact that the SM enters into a non-linear regime. The final contrast with ACAD-ROS+SM method is $10^{-5.33}$. At about 60 iterations, the ACAD-OSM algorithm also diverges. At this moment, it switches to the low-gain mode, allowing it to continue converging for 110 more iterations after which it diverge again at about 185 iterations. At this point we assume that we reached the limit of linearity for this matrix and we buid a new matrix, with an intermediary contrast of $10^{-6.35}$. After running our algorithm for eight interaction matrices, the final mean contrast is $10^{-9.41}$.

After showing in this section that ACAD-OSM shows better results both in contrast and in throughput at large separation than the previously developed active method, we show the impact of several aperture on the performance in the next section. 

\section{Impact of aperture discontinuities}
\label{sec:aperture_disco}

Here we compare the performance of this algorithm on 2 different aperture: WFIRST and an aperture currently under study for future space telescopes in the Segmented Coronagraph Design and Analysis (SCDA) program\footnote{This research program is led by NASA's Exoplanet Exploration Program (ExEP) \url{https://exoplanets.nasa.gov/system/internal_resources/details/original/211_SCDAApertureDocument050416.pdf}}. 

\begin{figure}
\begin{center}
 \includegraphics[trim= 1.5cm 0cm 0cm 0cm, clip = true,width = .48\textwidth]{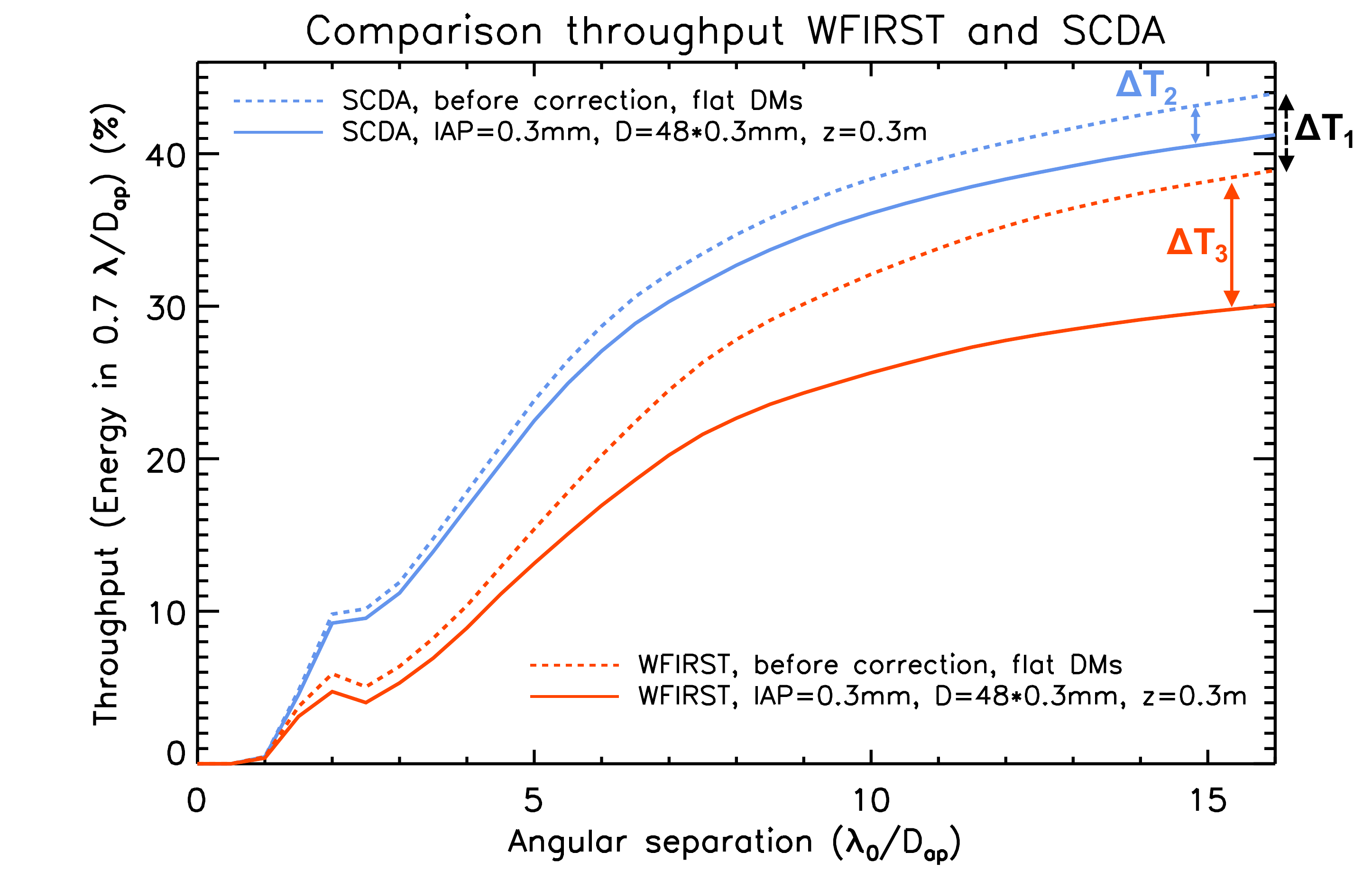}
  \includegraphics[trim= 1.5cm 0cm 0cm 0cm, clip = true,width = .48\textwidth]{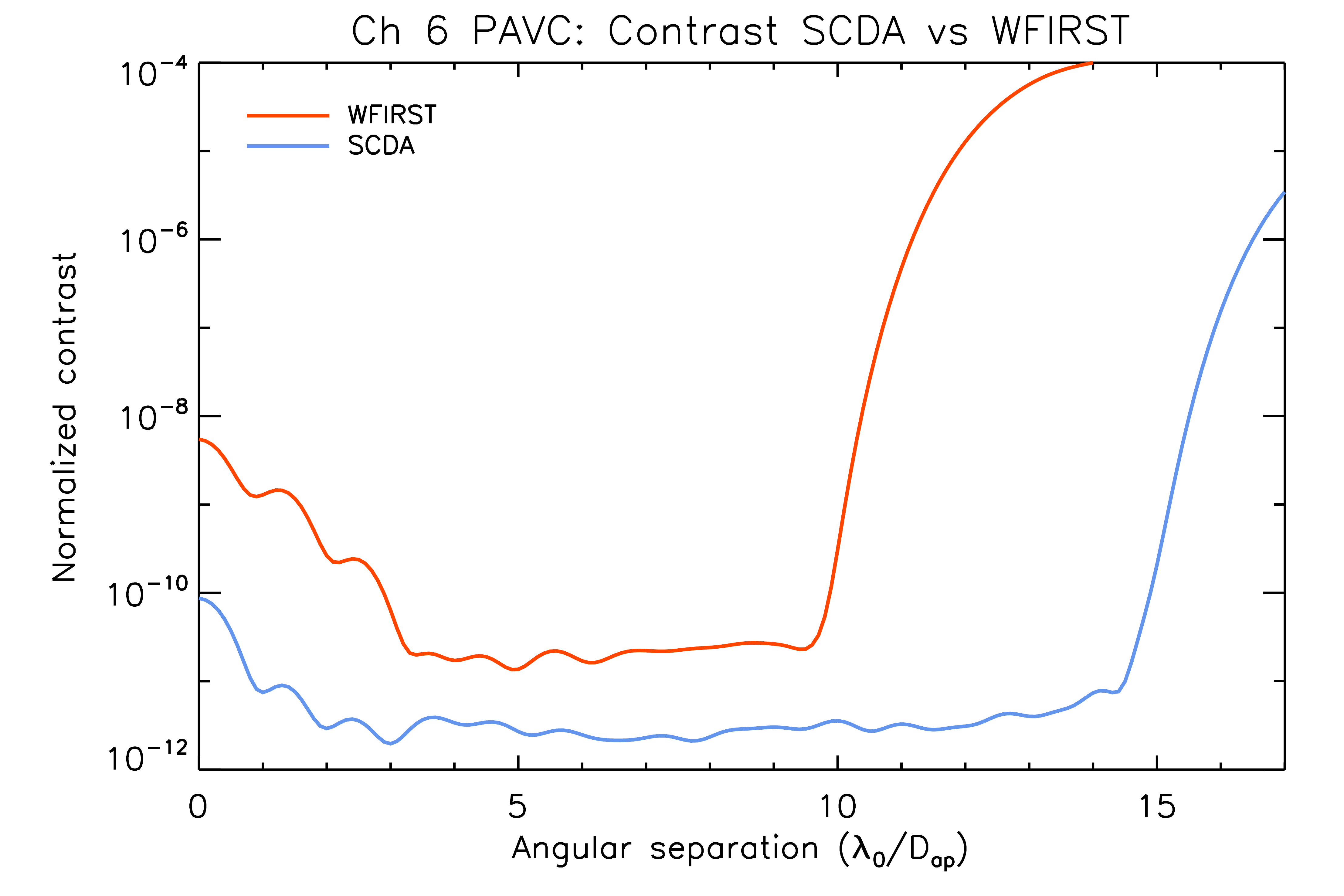}
  \end{center}
\caption[fig:wfirst_scda_throughput_compar_arrows]
{\label{fig:wfirst_scda_throughput_compar_arrows} Comparison WFIRST and SCDA performance for a charge 6 PAVC. The DM setup is $N_{act} = 48$, IAP = 0.3 mm, D = $48 * 0.3$ mm, $z = 0.3$ m and we use $\Delta \lambda /\lambda_0 = $ 10\% BW. In red, we show the results for the WFIRST aperture and in blue for the SCDA aperture. \textbf{Left:} Throughput performance. The dashed lines show the throughput with flat DMs, before any correction (due to the PAVC alone). The solid lines show the throughput after ACAD-OSM corrections. \textbf{Right:} contrast performance.}
\end{figure}

\subsection{Central obstruction}

Fig.~\ref{fig:wfirst_scda_throughput_compar_arrows} (left) shows the impact of the pupil discontinuities on off-axis throughput. We analyze the throughput before any correction (with flat DMS, dashed lines) and after the ACAD-OSM technique (solid lines) for the same DM setup and coronagraph (charge 6 PAVC, $N_{act} = 48$, IAP = 0.3 mm, D = $48 * 0.3$ mm, $z = 0.3$ m and we use $\Delta \lambda /\lambda_0 = $ 10\% BW), for the SCDA aperture (blue lines) and the WFIRST aperture (red lines). We notice the difference between the results in throughput for the two apertures before any correction is applied to the DM (difference between the dashed curves, $\Delta T_1$ = 5\% at 16 $\lambda_0/D_{ap}$). This difference is only related to the central obscuration size and to the method of the coronagraph apodization (PAVC here). Indeed, an increase in the central obscuration usually induces a decrease in the off-axis throughput \cite{ndiaye15b,fogarty17}.

\subsection{Struts and segmentation}

The influence of aperture discontinuities on the contrast level is mainly driven by the widths of the struts. For a given number of actuators, the SCDA aperture gives better contrast than the WFIRST aperture. For example, for the same coronagraph (PAVC) and DM setup ($N_{act} = 48$, IAP = 0.3 mm, D = $48 * 0.3$ mm, $z = 0.3$ m and we use $\Delta \lambda /\lambda_0 = $ 10\% BW), we obtain a result of $10^{-11.2}$ between 1.5 and 15 $\lambda_0/D_{ap}$ with the SCDA aperture and a contrast of $10^{-10.6}$ between 3 and 10 $\lambda_0/D_{ap}$ for the WFIRST aperture (Fig.~\ref{fig:wfirst_scda_throughput_compar_arrows}, right). 

For the same DM setup, the correction of large struts have a more important influence on the throughput, due to the larger strokes required on the DMs that can have a severe impact on the shape of the PSF at large separations. For the SCDA aperture, the difference in throughput between the initial state (Fig.~\ref{fig:wfirst_scda_throughput_compar_arrows}, left, blue dashed line) and final state (Fig.~\ref{fig:wfirst_scda_throughput_compar_arrows}, left, blue solid line) of the ACAD-OSM correction is $\Delta T_2$ = 3\% at 16 $\lambda_0/D_{ap}$. For the WFIRST aperture, the difference in throughput between the initial state (Fig.~\ref{fig:wfirst_scda_throughput_compar_arrows}, left, red dashed line) and final state (Fig.~\ref{fig:wfirst_scda_throughput_compar_arrows}, left, red solid line) of the ACAD-OSM correction is $\Delta T_3$ = 9\% at 16 $\lambda_0/D_{ap}$. 

Large struts have therefore not only an impact on the performance on contrast but also on the throughput of the correction, because of the high strokes they introduce on the DMs. The discontinuities due to the segmentation (only present in the SCDA aperture) have a less important impact on these metrics than the width of the struts. This is due to the the fact that segment discontinuities introduce less low-order spatial frequencies in the aperture, and therefore have a limited impact in the DH and can be corrected using less strokes on the DMs. However, segments in the aperture can fail, degrading the contrast in the DH. We study the correction of these segment failure in the next section.

\subsection{Missing segments in the aperture}
\label{sec:eelt}

\begin{figure}
\begin{center}
 \includegraphics[width = .48\textwidth, trim= 0.1cm 2cm 4.5cm 2cm, clip = true]{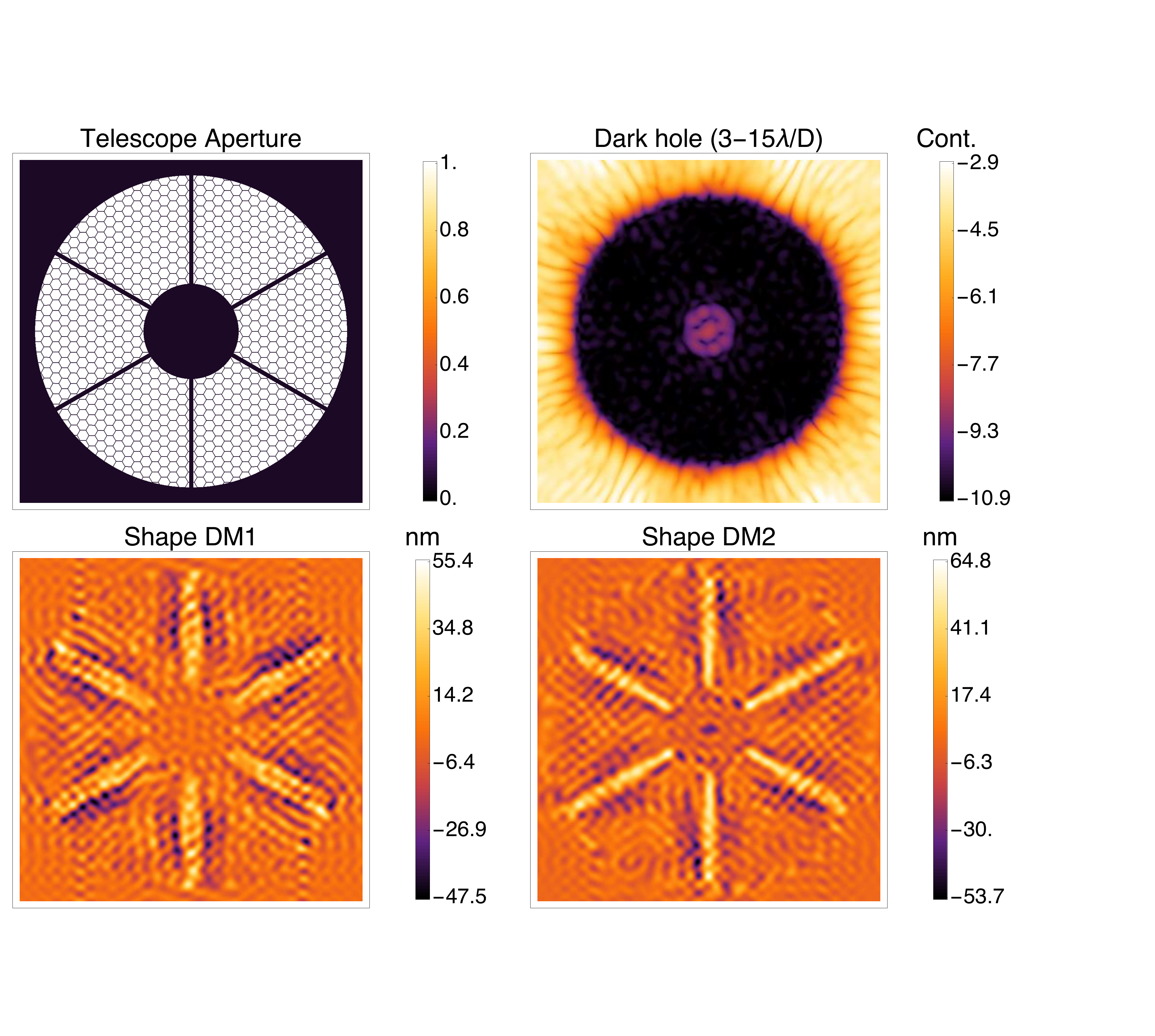}
 \includegraphics[width = .48\textwidth, trim= 0.1cm 2cm 4.5cm 2cm, clip = true]{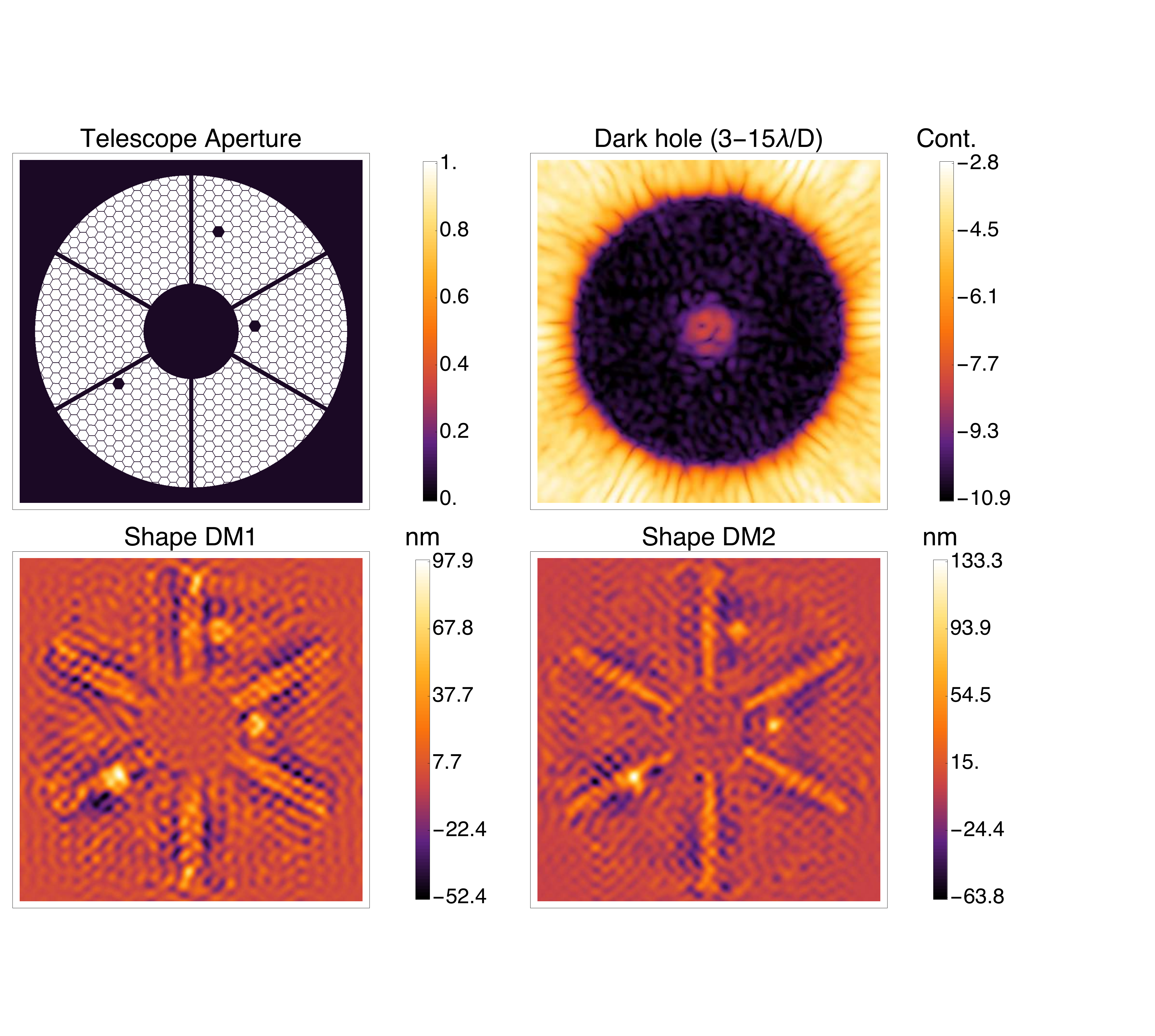}
 \end{center}
\caption[fig:eeltbs_dh]
{\label{fig:eelts_dh} \textbf{Left:} E-ELT aperture (charge 6 PAVC, $N_{act} = 48$, IAP = 0.3 mm, D = $48 * 0.3$ mm, $z = 0.3$ m, $\Delta \lambda /\lambda_0 = $ 10\%). Top left: E-ELT aperture. Top right: the final 3-15 $\lambda_0/D_{ap}$ DH. Bottom: the DM shapes obtained using ACAD-OSM for this solution. \textbf{Right:} Same aperture with  3 missing segments (charge 6 PAVC, $N_{act} = 48$, IAP = 0.3 mm, D = $48 * 0.3$ mm, $z = 0.3$ m, $\Delta \lambda /\lambda_0 = $ 10\%).}
\end{figure}

The aperture of the European Extremely Large Telescope (E-ELT, Figure~\ref{fig:eelts_dh}, top left) will be highly segmented (hundreds of mirrors). With so many segments, we can safely assume that a few of them will be inoperable or in maintenance every day. Active methods can quickly adapt to any change in the aperture geometry, compared with techniques involving static mirrors or apodization. In this section, we study the capabilities of ACAD-OSM in the presence of missing or inoperable segments.

\begin{figure}
\begin{center}
 \includegraphics[trim= 1.8cm 0.8cm 1cm 0.5cm, clip = true, width = .48\textwidth]{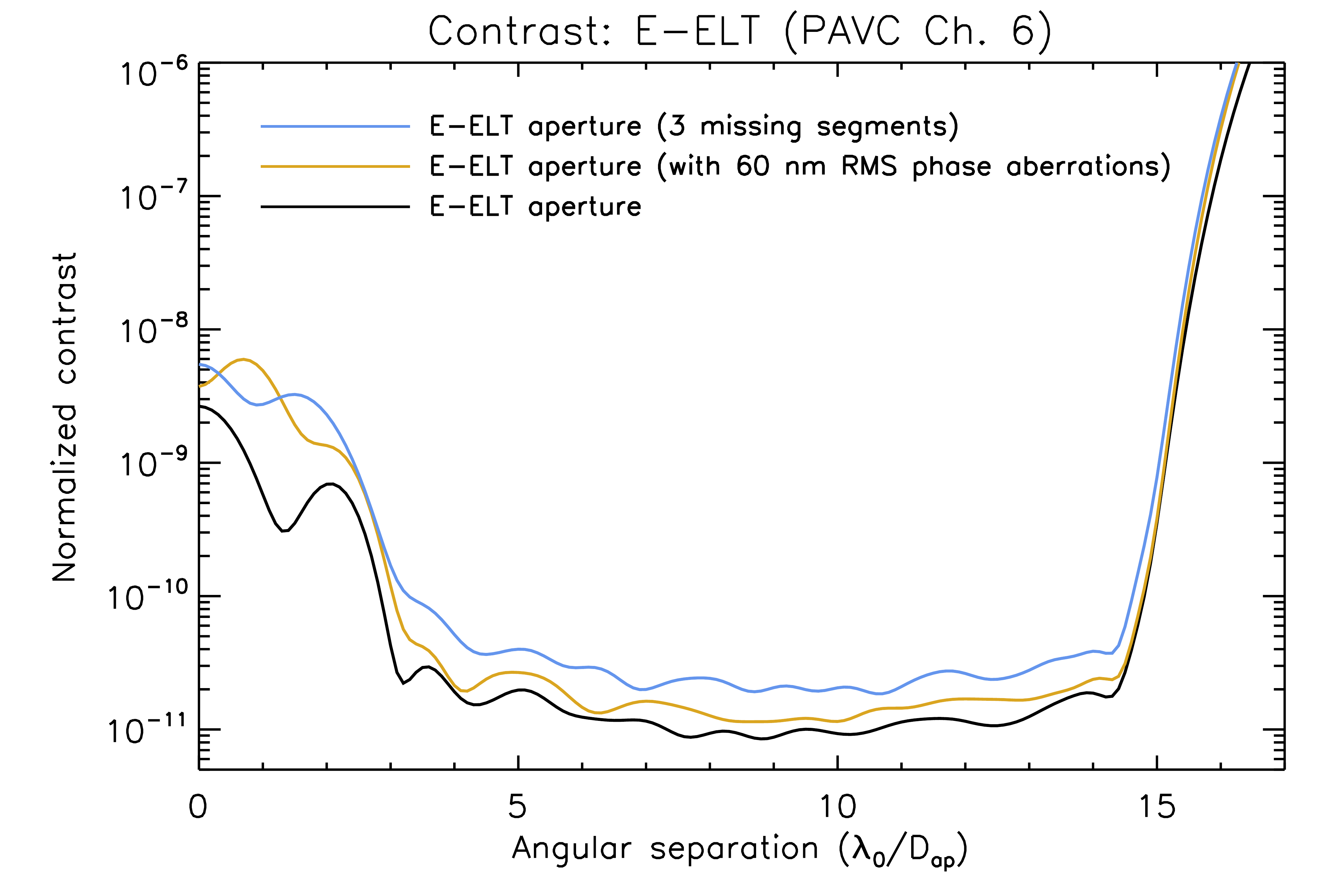}
 \includegraphics[trim= 1.5cm 1cm 1.0cm 0.5cm, clip = true,width = .48\textwidth]{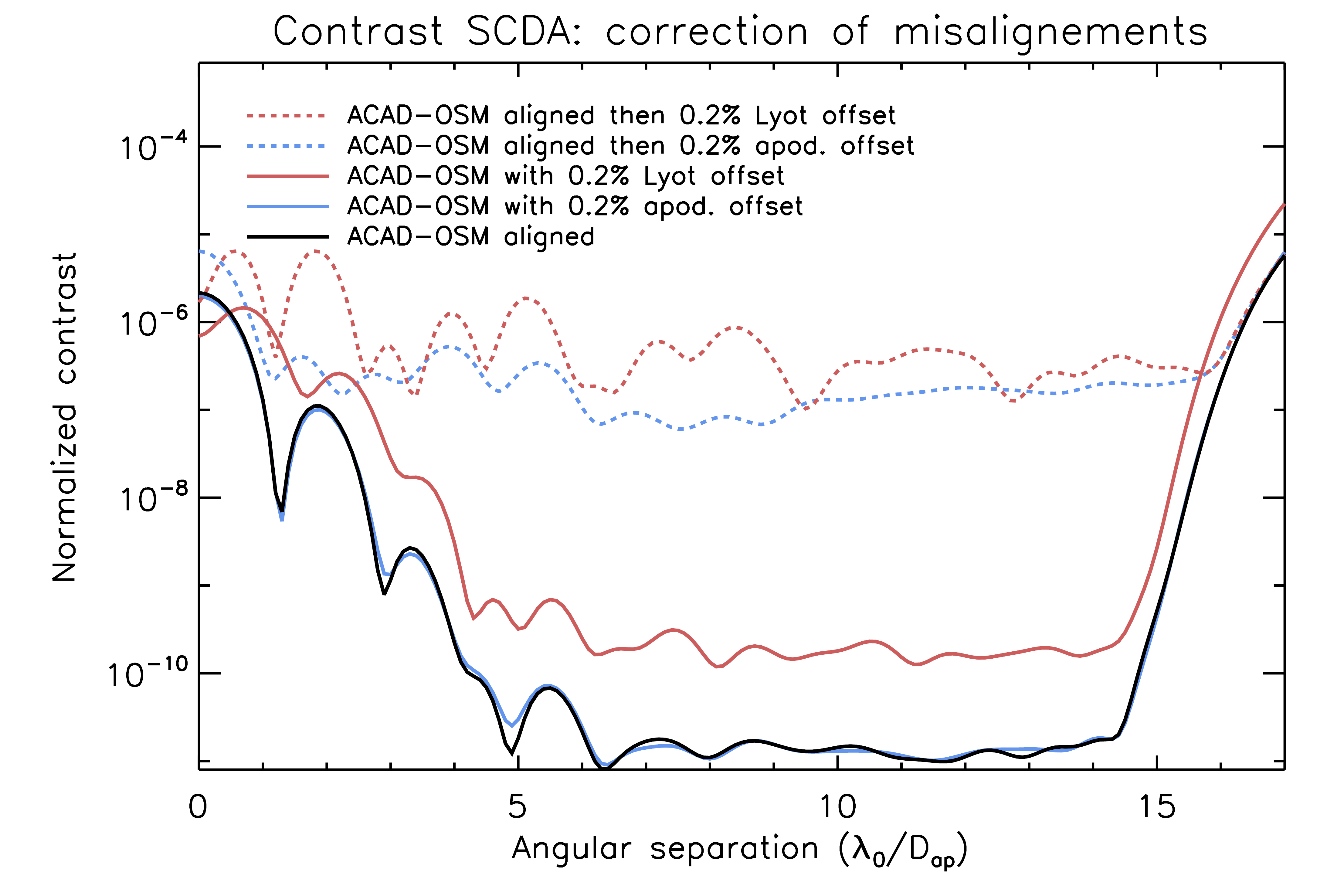}
  \end{center}
\caption[fig:eelt_and_misal_contrast]
{\label{fig:eelt_and_misal_contrast}\textbf{Left:} Contrast level for the E-ELT aperture (charge 6 PAVC, $N_{act} = 48$, IAP = 0.3 mm, D = $48 * 0.3$ mm, $z = 0.3$ m, $\Delta \lambda /\lambda_0 = $ 10\% BW). We show also the contrast levels for the same aperture with 3 missing segments (blue solid line) and in presence of a 60 nm RMS phase error (yellow line). \textbf{Right:} Contrast results for the SCDA aperture  with an APLC with and without mi-alignments. In black solid line is shown the results for an aligned APLC with an apodizer that is only optimized for the central obscuration in addition to a ACAD-OSM system. In yellow, the same system is used, but with a misalignment of 0.2\% in diameter of the apodization, corrected by the ACAD-OSM algorithm. In red, the same system is used, but with a misalignment of 0.2\% in diameter of the Lyot stop, corrected by the ACAD-OSM algorithm. The DM setup is $N_{act} = 48$, IAP = 0.3 mm, D = $48 * 0.3$ mm, $z = 0.3$ m, and the BW is $\Delta \lambda /\lambda_0 = $ 10\%.}
\end{figure}

We consider a Charge 6 PAVC with this aperture, using two $48 \times 48$ actuator DMs ($N_{act} = 48$, IAP = 0.3 mm, D = $48 * 0.3$ mm, $z = 0.3$ m), on a 10\% BW and on a 3-15 $\lambda_0/D_{ap}$ DH. The results for the E-ELT aperture are shown in Fig.~\ref{fig:eelts_dh} (left) in the absence of missing segment and in Fig~\ref{fig:eelts_dh} (right) with three missing segments. Fig.~\ref{fig:eelt_and_misal_contrast} (left) shows the contrast results in presence of missing segments (blue solid lines) in comparison with the nominal configuration with no missing segments (dark line). ACAD-OSM compensates for the image degradation caused by the missing segments, reaching within a factor of 3 of the nominal contrast inside the 3-15 $\lambda_0/D_{ap}$ DH. The throughput is barely impacted by these missing segments (not shown int his proceeding).

\section{Misalignements}
\label{sec:misalignements}

With hundreds of pixels in each of the apodizer directions compare to only tens of actuators, static apodizations control higher spatial frequencies and achieve larger and often deeper DH than active correction algorithms. However, the flexibility of an active method proves relevant to the evolution of the system with time, a situation for which static systems are not very robust. In Section~\ref{sec:eelt}, we proved that ACAD-OSM is robust to changes in the aperture. In this section, we study the impact of misalignments in the coronagraph system. 

\begin{figure}
\begin{center}
 \includegraphics[width = .48\textwidth, trim= 0.1cm 4.5cm 4.5cm 4cm, clip = true]{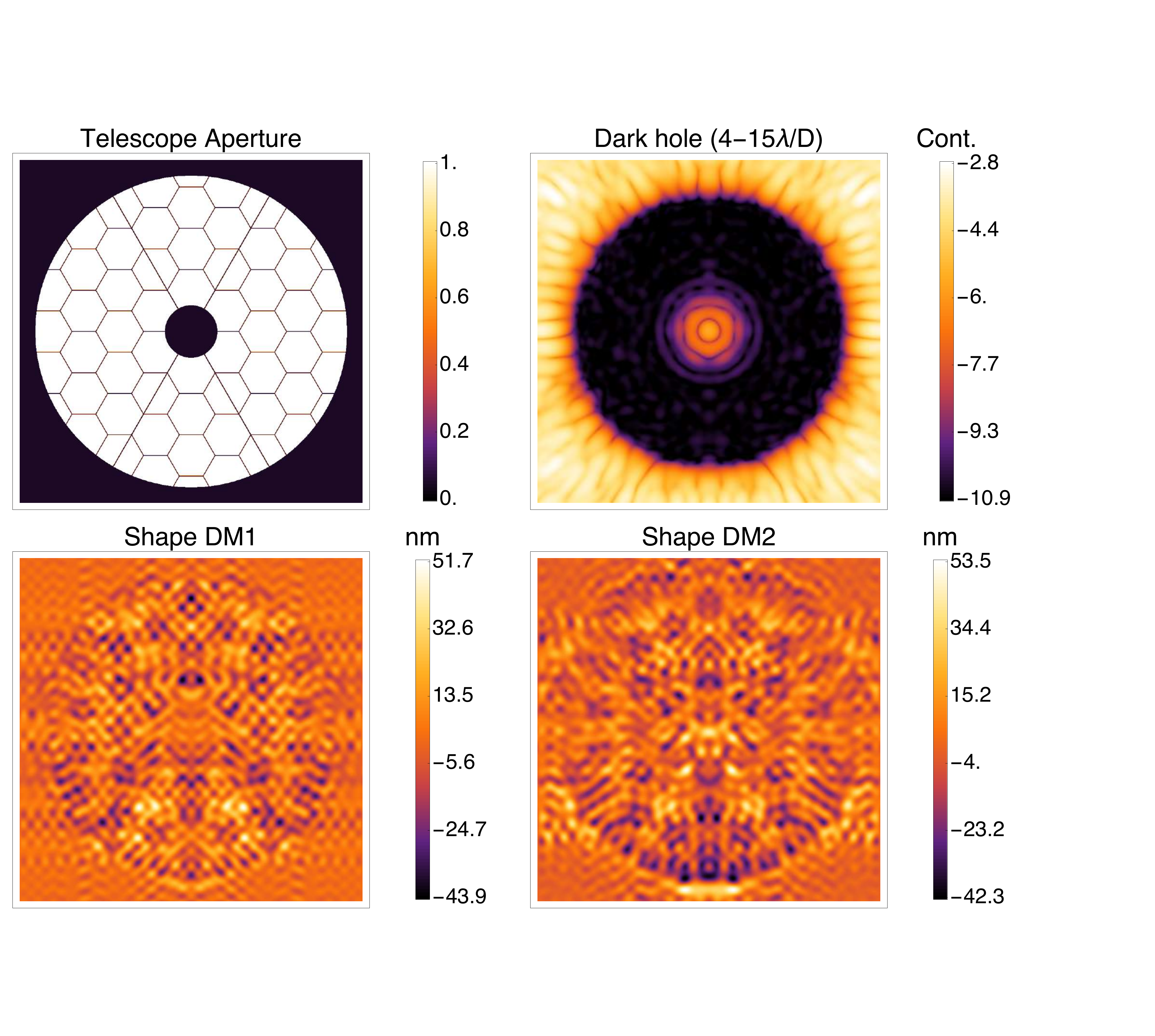}
 \includegraphics[width = .48\textwidth, trim= 0.1cm 4.5cm 4.5cm 4cm, clip = true]{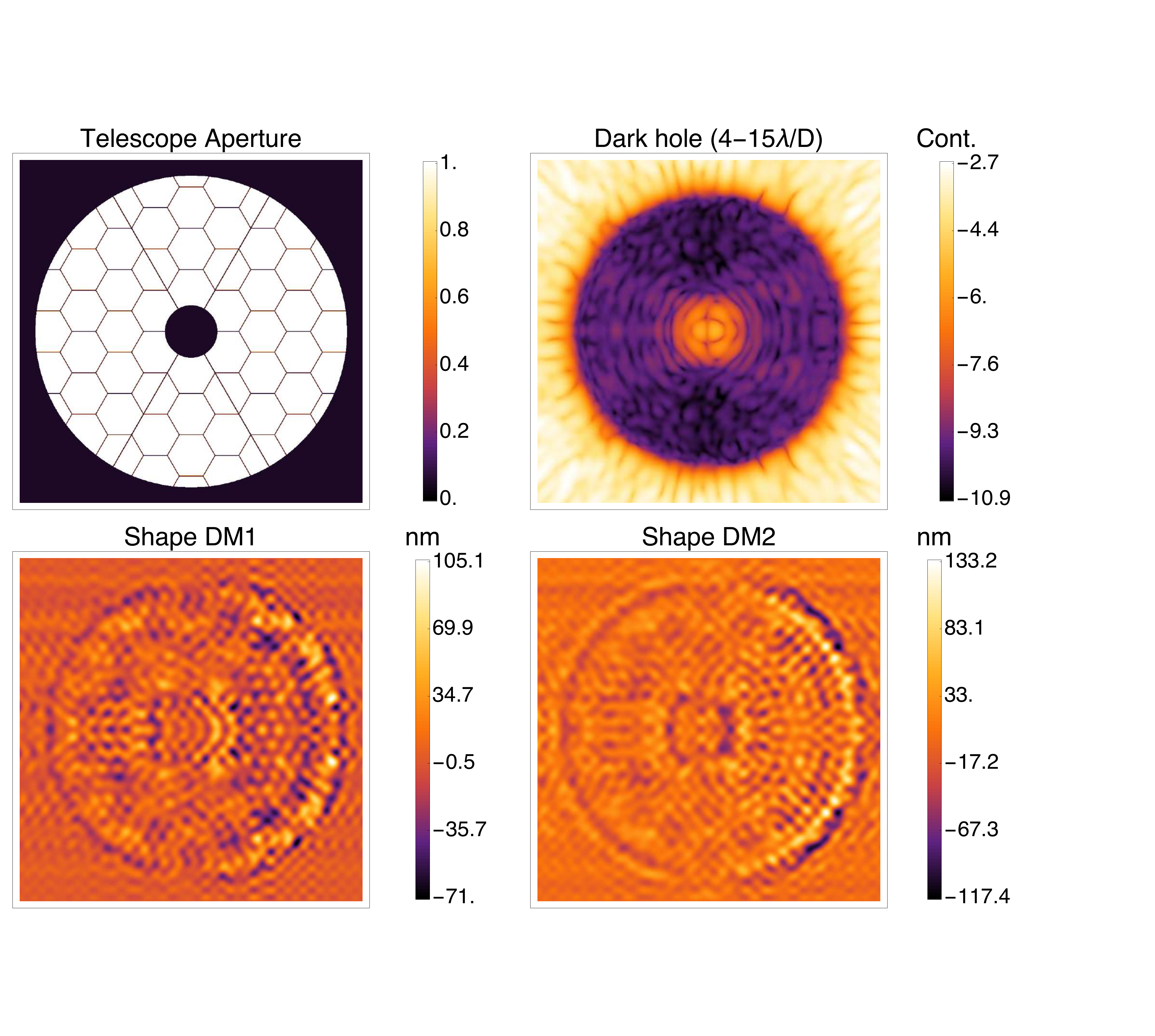}
 \end{center}
\caption[fig:misaldh]
{\label{fig:misaldh} Effect of apodization (left) and Lyot stop misalignments (right) on the DM shapes and final DH, for the SCDA aperture.}
\end{figure}

Several types of misalignment between the optics can degrade the contrast, e.g. a misalignment of the apodizer with respect to the re-imaged aperture in the entrance pupil plane. Here, we analyze the response of ACAD-OSM to a small misalignment in translation of the apodization in the system. For this test, an APLC with an apodizer optimized for a centrally obstructed pupil is used and misaligned by a 0.2\% $D_{ap}$ with respect to the optical axis. The results in Figure~\ref{fig:eelt_and_misal_contrast} (right) shows in black the initial results in contrast, after ACAD-OSM, when the system is aligned. The dashed blue line shows the degradation of contrast of this ACAD-OSM correction when a small misalignment of the apodization is introduced. We try to apply the ACAD-OSM system, this time in presence of an apodization misalignment. Starting the correction from the initial state (flat DMs) with this misalignment, the results obtained (blue solid curve) are similar to the ones obtained in the aligned case. This algorithm is therefore robust to a small misalignment of the apodizer in the pupil plane. We show the effects of apodization misalignment on the DM shapes and final DH in Figure \ref{fig:misaldh} (left). The misalignment only add a small asymmetry in the shapes of the DMs (in the bottom).

Finally, some apodized coronagraph designs can suffer from a misalignment of the Lyot stop with respect to the entrance pupil of the coronagraph. Two sorts of strategies have been developed for the design of the apodization. Some apodizations simply cancel out the energy from an on-axis star image inside the Lyot stop \cite{soummer03,mawet13,fogarty17}. Their corresponding coronagraphs are not sensitive to Lyot stop misalignments and a simple undersizing of the outer radius and oversizing of the inner radius can prevent this problem. However, other apodizations coherently recombines the electric field in the Lyot plane to produce a DH in the final image plane \cite{kasdin03,ruane15,ruane16,ndiaye15b, ndiaye16, zimmerman16}. Their corresponding coronagraphs are however very sensitive to a Lyot stop misalignment which breaks the fine recombination of coherent light in the relayed pupil plane and leads to a contrast degradation in the final image plane. The response of ACAD-OSM to this effect is studied with the APLC design where the Lyot stop is decentered by a 0.2\%$D_{ap}$ with respect to the optical axis. The effect of this misalignment are shown in Figure~\ref{fig:eelt_and_misal_contrast} (right). The correction is first made with a perfectly aligned system (black solid line) then a misalignment of the Lyot is introduced, degrading the contrast (dashed red line). Starting the correction from the initial state (flat DMs) with this Lyot misalignment, the results obtained (red solid curve) shows that ACAD-OSM can partially compensate for the loss in contrast. We show the effects of Lyot stop misalignment on the DM shapes and final DH in Figure \ref{fig:misaldh} (right). We see quickly that this effect (on the left on the DMs) is clearly the most difficult feature to correct by the DMs.

\section{Phase errors}
\label{sec:phase_errors}

We test the capabilities of this technique in the presence of both aperture discontinuities and realistic phase errors with for example low order aberrations caused by segments misalignments (Sec.~\ref{sec:phase_errors_segments}) or with higher order caused by optical aberrations (Sec.~\ref{sec:high_order_phase_errors}).

\subsection{Influence of small phase errors due to segment misalignments}
\label{sec:phase_errors_segments}

\begin{figure}
\begin{center}
 \includegraphics[width = .48\textwidth, trim= 0.1cm 4.5cm 4.5cm 4cm, clip = true]{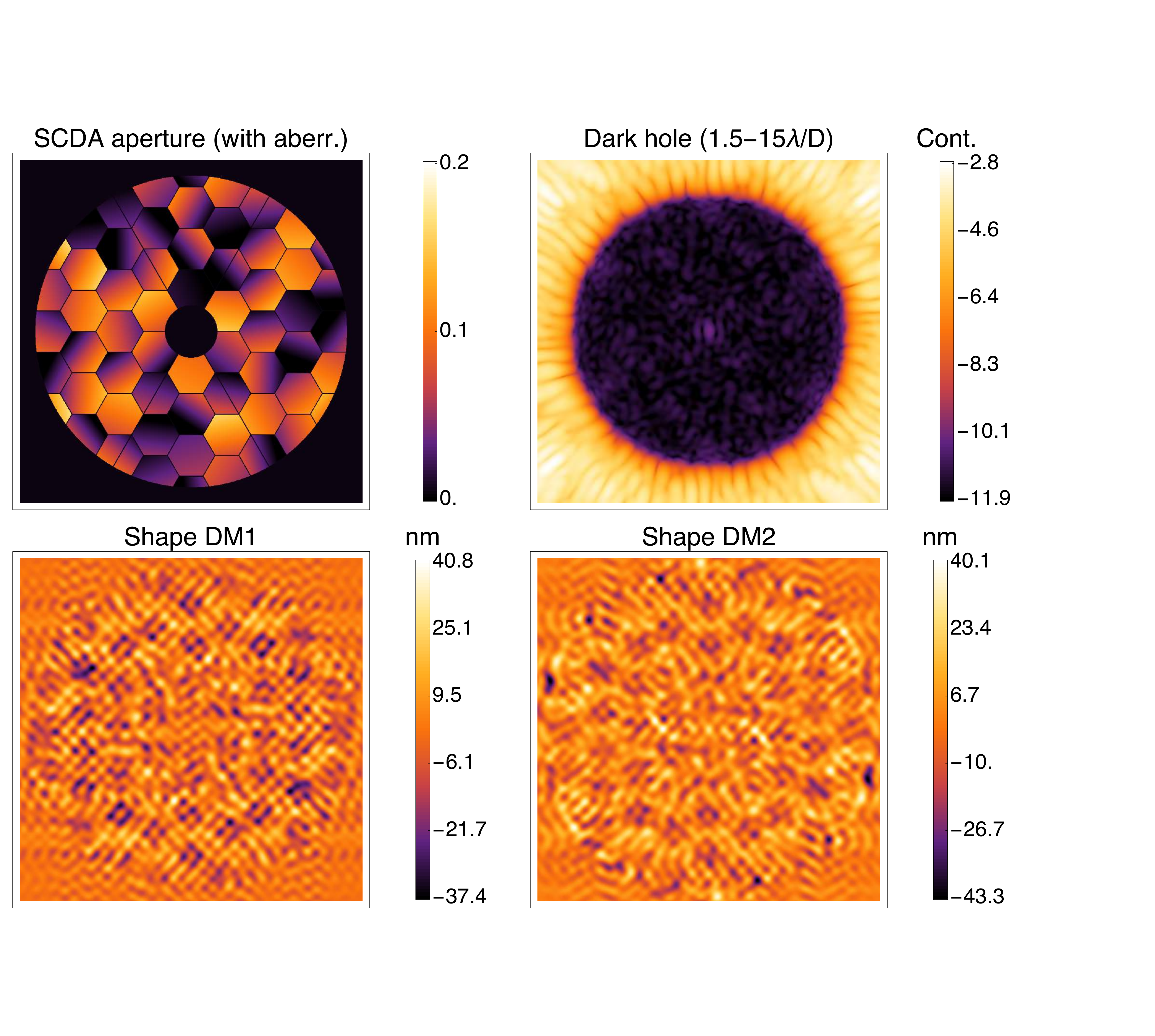}
 \includegraphics[width = .48\textwidth, trim= 0.1cm 4.5cm 4.5cm 4cm, clip = true]{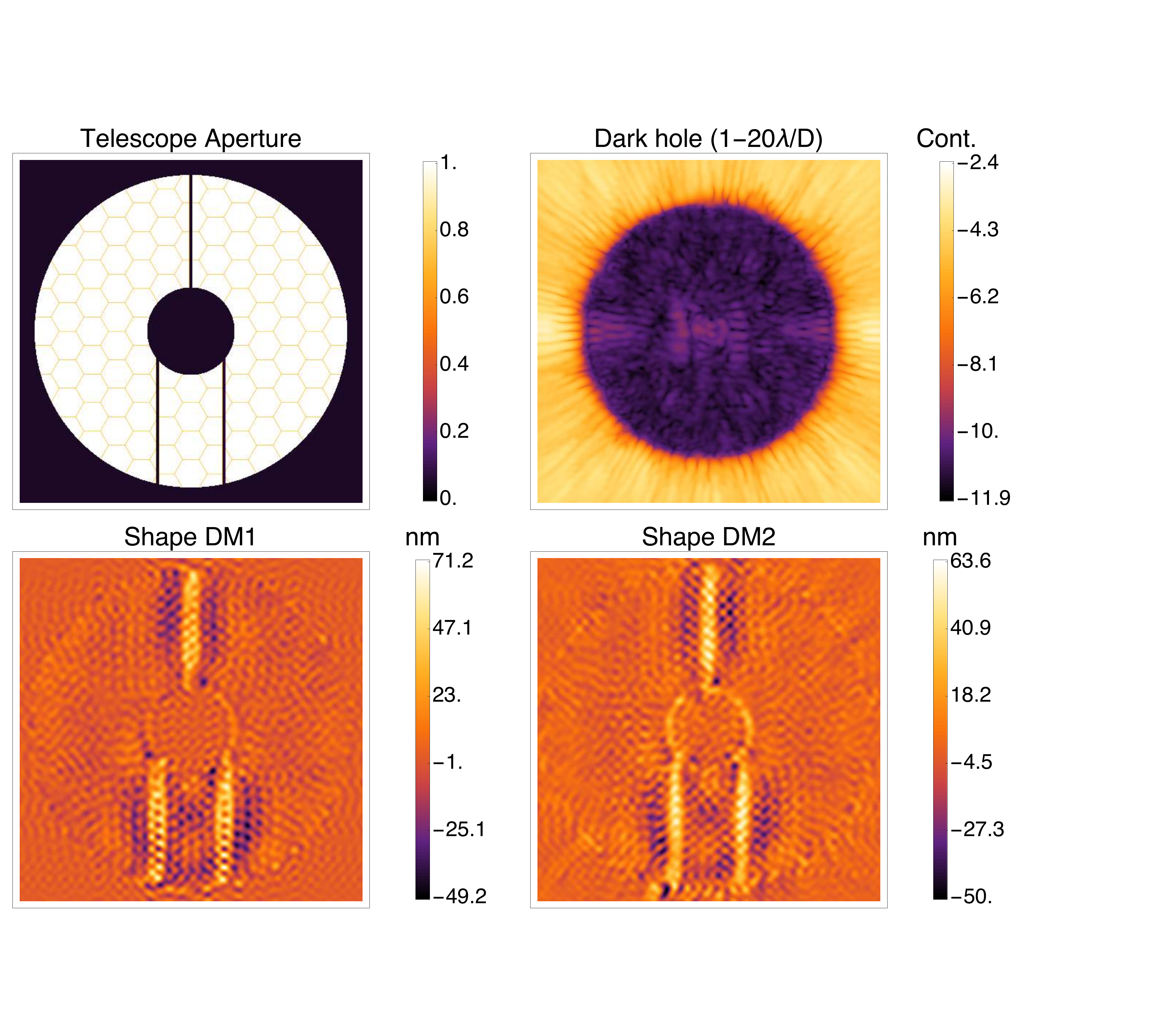}
 \end{center}
\caption[fig:scda_dh_segmentphase_and_luvoir]
{\label{fig:scda_dh_segmentphase_and_luvoir} \textbf{Left:} SCDA aperture (Charge 6 PAVC coronagraph, $N_{act} = 48$, IAP = 0.3 mm, D = $48 * 0.3$ mm, $z = 0.3$ m) with a 10\% BW in the presence of phase errors due to segment mis-alignements. Top left: SCDA aperture with phase aberrations. Top right: the final 1.5-15 $\lambda_0/D_{ap}$ DH. Bottom: the DM shapes obtained using ACAD-OSM for this solution. \textbf{Right}:  LUVOIR aperture (Charge 6 PAVC coronagraph, $N_{act} = 64$, IAP = 0.5 mm, D = $48 * 0.3$ mm, $z = 0.5$ m) with a 20\% BW. Top left: LUVOIR aperture with phase aberrations. Top right: the final 1-20 $\lambda_0/D_{ap}$ DH. Bottom: the DM shapes obtained using ACAD-OSM for this solution.}
\end{figure}

All the examples so far have assumed a perfect cophasing of all the segments and the aperture does not include phase aberrations. We here show that a small amount of phase due to imperfect alignment of the segments in piston (10 nm peak-to-valley) and tip-tilt (10 nm peak-to-valley) does not limit our method. Figure~\ref{fig:scda_dh_segmentphase_and_luvoir} (left) shows a DH with our setup, in the presence of phase errors due to segment misalignments, after the correction with ACAD-OSM. The contrast and throughput (not shown here) are almost identical with these small phase aberrations and in the absence of aberrations.

\subsection{Influence of high-order spatial frequency phase errors}
\label{sec:high_order_phase_errors}

Since ACAD-OSM uses the focal plane contrast to drive the correction, it addresses aperture discontinuity effects in the focal plane exactly like the effects of continuous phase or amplitude apertures. However, one can wonder if the correction can be limited in the presence of important phase aberrations creating speckles, that can partially masks the effect of the aperture discontinuities. To study this phenomenon, we introduce a large (60 nm) phase error in the entrance pupil plane with the E-ELT aperture and attempt to correct for both this phase and the aperture discontinuities simultaneously. The phase error has an amplitude of 60 nm RMS, with a power spectral density law of $f^{-3}$, where $f$ denotes the spatial frequency. The results in contrast are shown with a yellow curve in Fig.~\ref{fig:eelt_and_misal_contrast} (left). It shows no significant difference with the case of the E-ELT aperture wihtout phase. We also observe similar throughput levels (not shown here).

\section{Application to the LUVOIR aperture}
\label{sec:LUVOIR}

In this section, we present the result obtained on the current LUVOIR aperture. We show the DM shapes and the obtained 1-20 $\lambda_0/D_{ap}$ DH in Fig.~\ref{fig:scda_dh_segmentphase_and_luvoir} (right). We used a Charge 6 PAVC coronagraph, $N_{act} = 64$, IAP = 0.3 mm, D = $48 * 0.3$ mm, $z = 0.5$ m) with a 20\% BW. The mean contrast obtained was $10^{-10.6}$ in the 1-20 $\lambda_0/D_{ap}$ DH. The mean throughput in the DH is 35\%.

\section{Conclusion}

In this proceeding, we presented the active correction of aperture discontinuities - optimized stroke minimization, ACAD-OSM, a new active method to correct for the discontinuities in the aperture. This method uses an adaptive interaction matrix to control two deformable mirrors. In the first part, we described the algorithm and its superiority over previous active methods \cite{pueyo13}. We have illustrated its capabilities with some specific cases (WFIRST, E-ELT, SCDA aperture and LUVOIR). As the main results of this proceeding, (1) the ACAD-OSM method applies to any coronagraph or aperture and reaches the symbolic $10^{-10}$ exo-Earth contrast level with a 10\% bandwidth and existing DM configurations; (2) this active technique handles --- without any revision to the algorithm --- changing or unknown optical aberrations or discontinuities in the pupil, including optical design misalignments, missing segments and phase errors. The experimental validation of this technique should soon take place on the High-Contrast Imager for Complex Aperture Telescopes (HiCAT) optical test bench \cite{ndiaye15} located at the Space Telescope Science Institute.

We conclude that a static optical design (apodization or static mirrors) alone is not the preferable approach to compensate for aperture discontinuities. We instead advocate for an active technique with two-DM correction, due to its adaptability to evolving or unknown optical design, such as coronagraph optic misalignments or static phase errors. However, we acknowledge that the current generation of DMs does not offer enough degrees of freedom to correct by itself for all the features of a complex aperture. In particular, the correction of the central obscuration with DMs only, is for the moment out of reach since it involves very large strokes. A solution combining the advantages of both methods (static and active correction) should be pursued for future coronagraphic instruments.

\acknowledgments     
 
This material is based upon work partially carried out under subcontract \#1496556 with the Jet Propulsion Laboratory funded by NASA and administered by the California Institute of Technology.


\bibliography{biblio_spie_17}   

\begin{thebibliography}{10}

\bibitem{macintosh06}
Macintosh, B., Troy, M., Doyon, R., Graham, J., Baker, K., Bauman, B., Marois,
  C., Palmer, D., Phillion, D., Poyneer, L., Crossfield, I., Dumont, P.,
  Levine, B.~M., Shao, M., Serabyn, G., Shelton, C., Vasisht, G., Wallace,
  J.~K., Lavigne, J.-F., Valee, P., Rowlands, N., Tam, K., and Hackett, D.,
  ``Extreme adaptive optics for the {Thirty} {Meter} {Telescope},'' in [{\em
  Proceedings of the {SPIE}}{\nolinebreak\hspace{0.1em}]},  {\em Proceedings of
  the {SPIE}} {\bf 6272},  62720N (June 2006).

\bibitem{kasper08}
Kasper, M.~E., Beuzit, J.-L., Verinaud, C., Yaitskova, N., Baudoz, P.,
  Boccaletti, A., Gratton, R.~G., Hubin, N., Kerber, F., Roelfsema, R., Schmid,
  H.~M., Thatte, N.~A., Dohlen, K., Feldt, M., Venema, L., and Wolf, S.,
  ``{EPICS}: the exoplanet imager for the {E}-{ELT},'' in [{\em Proceedings of
  the {SPIE}}{\nolinebreak\hspace{0.1em}]},  {\em Proceedings of the {SPIE}}
  {\bf 7015},  46 (July 2008).

\bibitem{davies10}
Davies, R., Ageorges, N., Barl, L., Bedin, L.~R., Bender, R., Bernardi, P.,
  Chapron, F., Clenet, Y., Deep, A., Deul, E., Drost, M., Eisenhauer, F.,
  Falomo, R., Fiorentino, G., Förster~Schreiber, N.~M., Gendron, E., Genzel,
  R., Gratadour, D., Greggio, L., Grupp, F., Held, E., Herbst, T., Hess, H.-J.,
  Hubert, Z., Jahnke, K., Kuijken, K., Lutz, D., Magrin, D., Muschielok, B.,
  Navarro, R., Noyola, E., Paumard, T., Piotto, G., Ragazzoni, R., Renzini, A.,
  Rousset, G., Rix, H.-W., Saglia, R., Tacconi, L., Thiel, M., Tolstoy, E.,
  Trippe, S., Tromp, N., Valentijn, E.~A., Verdoes~Kleijn, G., and Wegner, M.,
  ``{MICADO}: the {E}-{ELT} adaptive optics imaging camera,'' in [{\em
  Proceedings of the {SPIE}}{\nolinebreak\hspace{0.1em}]},  {\em Proceedings of
  the {SPIE}} {\bf 7735},  77352A (July 2010).

\bibitem{quanz15c}
Quanz, S.~P., Crossfield, I., Meyer, M.~R., Schmalzl, E., and Held, J.,
  ``Direct detection of exoplanets in the 3-10 μm range with
  {E}-{ELT}/{METIS},'' {\em International Journal of Astrobiology}~{\bf 14},
  279--289 (Apr. 2015).

\bibitem{spergel15}
Spergel, D., Gehrels, N., Baltay, C., Bennett, D., Breckinridge, J., Donahue,
  M., Dressler, A., Gaudi, B.~S., Greene, T., Guyon, O., Hirata, C., Kalirai,
  J., Kasdin, N.~J., Macintosh, B., Moos, W., Perlmutter, S., Postman, M.,
  Rauscher, B., Rhodes, J., Wang, Y., Weinberg, D., Benford, D., Hudson, M.,
  Jeong, W.-S., Mellier, Y., Traub, W., Yamada, T., Capak, P., Colbert, J.,
  Masters, D., Penny, M., Savransky, D., Stern, D., Zimmerman, N., Barry, R.,
  Bartusek, L., Carpenter, K., Cheng, E., Content, D., Dekens, F., Demers, R.,
  Grady, K., Jackson, C., Kuan, G., Kruk, J., Melton, M., Nemati, B., Parvin,
  B., Poberezhskiy, I., Peddie, C., Ruffa, J., Wallace, J.~K., Whipple, A.,
  Wollack, E., and Zhao, F., ``Wide-{Field} {InfrarRed} {Survey}
  {Telescope}-{Astrophysics} {Focused} {Telescope} {Assets} {WFIRST}-{AFTA}
  2015 {Report},'' {\em arXiv:1503.03757 [astro-ph]}  (Mar. 2015).
\newblock arXiv: 1503.03757.

\bibitem{dalcanton15}
Dalcanton, J., Seager, S., Aigrain, S., Battel, S., Brandt, N., Conroy, C.,
  Feinberg, L., Gezari, S., Guyon, O., Harris, W., Hirata, C., Mather, J.,
  Postman, M., Redding, D., Schiminovich, D., Stahl, H.~P., and Tumlinson, J.,
  ``From {Cosmic} {Birth} to {Living} {Earths}: {The} {Future} of {UVOIR}
  {Space} {Astronomy},'' {\em arXiv:1507.04779 [astro-ph]}  (July 2015).
\newblock arXiv: 1507.04779.

\bibitem{mennesson16}
{Mennesson}, B., {Gaudi}, S., {Seager}, S., {Cahoy}, K., {Domagal-Goldman}, S.,
  {Feinberg}, L., {Guyon}, O., {Kasdin}, J., {Marois}, C., {Mawet}, D.,
  {Tamura}, M., {Mouillet}, D., {Prusti}, T., {Quirrenbach}, A., {Robinson},
  T., {Rogers}, L., {Scowen}, P., {Somerville}, R., {Stapelfeldt}, K., {Stern},
  D., {Still}, M., {Turnbull}, M., {Booth}, J., {Kiessling}, A., {Kuan}, G.,
  and {Warfield}, K., ``{The Habitable Exoplanet (HabEx) Imaging Mission:
  preliminary science drivers and technical requirements},'' in [{\em Space
  Telescopes and Instrumentation 2016: Optical, Infrared, and Millimeter
  Wave}{\nolinebreak\hspace{0.1em}]},  {\em \procspie} {\bf 9904},  99040L
  (July 2016).

\bibitem{pueyo13}
Pueyo, L. and Norman, C., ``High-contrast {Imaging} with an {Arbitrary}
  {Aperture}: {Active} {Compensation} of {Aperture} {Discontinuities},'' {\em
  The Astrophysical Journal}~{\bf 769},  102 (June 2013).

\bibitem{fogarty17}
{Fogarty}, K., {Pueyo}, L., {Mazoyer}, J., and {N'Diaye}, M., ``{Polynomial
  Apodizers for Centrally Obscured Vortex Coronagraphs},'' {\em ArXiv e-prints
  (submitted to ApJ)}  (Mar. 2017).

\bibitem{mazoyer15}
Mazoyer, J., Pueyo, L., Norman, C., N'Diaye, M., Mawet, D., Soummer, R.,
  Perrin, M., Choquet, Ã., and Carlotti, A., ``Active correction of aperture
  discontinuities ({ACAD}) for space telescope pupils: a parametic analysis,''
  in [{\em Proceedings of the {SPIE}}{\nolinebreak\hspace{0.1em}]},  {\em
  Proceedings of the {SPIE}} {\bf 9605},  96050M (Sept. 2015).

\bibitem{soummer03}
Soummer, R., Aime, C., and Falloon, P.~E., ``Stellar coronagraphy with prolate
  apodized circular apertures,'' {\em Astronomy and Astrophysics}~{\bf 397},
  1161--1172 (Jan. 2003).

\bibitem{soummer11}
Soummer, R., Sivaramakrishnan, A., Pueyo, L., Macintosh, B., and Oppenheimer,
  B.~R., ``Apodized {Pupil} {Lyot} {Coronagraphs} for {Arbitrary} {Apertures}.
  {III}. {Quasi}-achromatic {Solutions},'' {\em The Astrophysical Journal}~{\bf
  729},  144 (Mar. 2011).

\bibitem{ndiaye16}
N'Diaye, M., Soummer, R., Pueyo, L., Carlotti, A., Stark, C.~C., and Perrin,
  M.~D., ``Apodized {Pupil} {Lyot} {Coronagraphs} for {Arbitrary} {Apertures}.
  {V}. {Hybrid} {Shaped} {Pupil} {Designs} for {Imaging} {Earth}-like planets
  with {Future} {Space} {Observatories},'' {\em The Astrophysical Journal}~{\bf
  818},  163 (Feb. 2016).

\bibitem{pueyo14}
Pueyo, L., Norman, C., Soummer, R., Perrin, M., N'Diaye, M., Choquet, Ã.,
  Hoffmann, J., Carlotti, A., and Mawet, D., ``High contrast imaging with an
  arbitrary aperture: active correction of aperture discontinuities:
  fundamental limits and practical trade- offs,'' in [{\em Proceedings of the
  {SPIE}}{\nolinebreak\hspace{0.1em}]},  {\em Proceedings of the {SPIE}} {\bf
  9143},  914321 (Aug. 2014).

\bibitem{mazoyer16b}
Mazoyer, J., Pueyo, L., Norman, C., N'Diaye, M., van~der Marel, R.~P., and
  Soummer, R., ``Active compensation of aperture discontinuities for
  {WFIRST}-{AFTA}: analytical and numerical comparison of propagation methods
  and preliminary results with a {WFIRST}-{AFTA}-like pupil,'' {\em Journal of
  Astronomical Telescopes, Instruments, and Systems}~{\bf 2},  011008 (Mar.
  2016).

\bibitem{pueyo09}
Pueyo, L., Kay, J., Kasdin, N.~J., Groff, T., McElwain, M., Give'on, A., and
  Belikov, R., ``Optimal dark hole generation via two deformable mirrors with
  stroke minimization,'' {\em Applied Optics}~{\bf 48},  6296 (Nov. 2009).

\bibitem{guyon2005}
Guyon, O., Pluzhnik, E.~A., Galicher, R., Martinache, F., Ridgway, S.~T., and
  Woodruff, R.~A., ``Exoplanet {Imaging} with a {Phase}-induced {Amplitude}
  {Apodization} {Coronagraph}. {I}. {Principle},'' {\em The Astrophysical
  Journal}~{\bf 622},  744--758 (Mar. 2005).

\bibitem{mazoyer13c}
Mazoyer, J., Baudoz, P., Galicher, R., Mas, M., and Rousset, G., ``Estimation
  and correction of wavefront aberrations using the self-coherent camera:
  laboratory results,'' {\em Astronomy and Astrophysics}~{\bf 557},  9 (Sept.
  2013).

\bibitem{baudoz06}
Baudoz, P., Boccaletti, A., Baudrand, J., and Rouan, D., ``The
  {Self}-{Coherent} {Camera}: a new tool for planet detection,'' in [{\em
  Proceedings of the {IAU} {Colloquium}}{\nolinebreak\hspace{0.1em}]},  {\em
  Proceedings of the {IAU} {Colloquium}},  553--558 (2006).

\bibitem{ndiaye16b}
N'Diaye, M., Vigan, A., Dohlen, K., Sauvage, J.-F., Caillat, A., Costille, A.,
  Girard, J. H.~V., Beuzit, J.-L., Fusco, T., Blanchard, P., Le~Merrer, J.,
  Le~Mignant, D., Madec, F., Moreaux, G., Mouillet, D., Puget, P., and Zins,
  G., ``Calibration of quasi-static aberrations in exoplanet direct-imaging
  instruments with a {Zernike} phase-mask sensor. {II}. {Concept} validation
  with {ZELDA} on {VLT}/{SPHERE},'' {\em Astronomy and Astrophysics}~{\bf 592},
   A79 (Aug. 2016).

\bibitem{borde06}
Bord\'e, P.~J. and Traub, W.~A., ``High-{Contrast} {Imaging} from {Space}:
  {Speckle} {Nulling} in a {Low}-{Aberration} {Regime},'' {\em The
  Astrophysical Journal}~{\bf 638},  488 (Feb. 2006).

\bibitem{giveon07}
Give'On, A., Belikov, R., Shaklan, S., and Kasdin, J., ``Closed loop, {DM}
  diversity-based, wavefront correction algorithm for high contrast imaging
  systems,'' {\em Optics Express}~{\bf 15}(19),  12338--12343 (2007).

\bibitem{paul13}
Paul, B., Sauvage, J.-F., and Mugnier, L.~M., ``Coronagraphic phase diversity:
  performance study and laboratory demonstration,'' {\em Astronomy and
  Astrophysics}~{\bf 552},  48 (Apr. 2013).

\bibitem{riggs16}
Riggs, A. J.~E., Kasdin, N.~J., and Groff, T.~D., ``Recursive starlight and
  bias estimation for high-contrast imaging with an extended {Kalman} filter,''
  {\em Journal of Astronomical Telescopes, Instruments, and Systems}~{\bf 2},
  011017 (Jan. 2016).

\bibitem{trauger07}
Trauger, J.~T. and Traub, W.~A., ``A laboratory demonstration of the capability
  to image an {Earth}-like extrasolar planet,'' {\em Nature}~{\bf 446},
  771--773 (Apr. 2007).

\bibitem{mazoyer14}
Mazoyer, J., Baudoz, P., Galicher, R., and Rousset, G., ``High-contrast imaging
  in polychromatic light with the self-coherent camera,'' {\em Astronomy and
  Astrophysics}~{\bf 564},  L1 (Apr. 2014).

\bibitem{ndiaye15b}
N'Diaye, M., Pueyo, L., and Soummer, R., ``Apodized {Pupil} {Lyot}
  {Coronagraphs} for {Arbitrary} {Apertures}. {IV}. {Reduced} {Inner} {Working}
  {Angle} and {Increased} {Robustness} to {Low}-order {Aberrations},'' {\em The
  Astrophysical Journal}~{\bf 799},  225 (Feb. 2015).

\bibitem{mawet13}
Mawet, D., Pueyo, L., Carlotti, A., Mennesson, B., Serabyn, E., and Wallace,
  J.~K., ``Ring-apodized {Vortex} {Coronagraphs} for {Obscured} {Telescopes}.
  {I}. {Transmissive} {Ring} {Apodizers},'' {\em The Astrophysical Journal
  Supplement Series}~{\bf 209},  7 (Nov. 2013).

\bibitem{kasdin03}
{Kasdin}, N.~J., {Vanderbei}, R.~J., {Spergel}, D.~N., and {Littman}, M.~G.,
  ``{Extrasolar Planet Finding via Optimal Apodized-Pupil and Shaped-Pupil
  Coronagraphs},'' {\em \apj}~{\bf 582},  1147--1161 (Jan. 2003).

\bibitem{ruane15}
Ruane, G.~J., Huby, E., Absil, O., Mawet, D., Delacroix, C., Carlomagno, B.,
  and Swartzlander, G.~A., ``Lyot-plane phase masks for improved high-contrast
  imaging with a vortex coronagraph,'' {\em Astronomy and Astrophysics}~{\bf
  583},  A81 (Nov. 2015).

\bibitem{ruane16}
Ruane, G., Jewell, J., Mawet, D., Pueyo, L., and Shaklan, S., ``Apodized vortex
  coronagraph designs for segmented aperture telescopes,'' in [{\em Proceedings
  of the {SPIE}}{\nolinebreak\hspace{0.1em}]},  {\em Proceedings of the {SPIE}}
  {\bf 9912},  99122L--99122L--13 (2016).

\bibitem{zimmerman16}
{Zimmerman}, N.~T., {Eldorado Riggs}, A.~J., {Jeremy Kasdin}, N., {Carlotti},
  A., and {Vanderbei}, R.~J., ``{Shaped pupil Lyot coronagraphs: high-contrast
  solutions for restricted focal planes},'' {\em Journal of Astronomical
  Telescopes, Instruments, and Systems}~{\bf 2},  011012 (Jan. 2016).

\bibitem{ndiaye15}
N'Diaye, M., Mazoyer, J., Choquet, Ã., Pueyo, L., Perrin, M.~D., Egron, S.,
  Leboulleux, L., Levecq, O., Carlotti, A., Long, C.~A., Lajoie, R., and
  Soummer, R., ``High-contrast imager for complex aperture telescopes
  ({HiCAT}): 3. first lab results with wavefront control,'' in [{\em
  Proceedings of the {SPIE}}{\nolinebreak\hspace{0.1em}]},  {\em Proceedings of
  the {SPIE}} {\bf 9605},  96050I (Sept. 2015).

\end{thebibliography}
\bibliographystyle{spiebib}   

\end{document}